\documentclass{emulateapj}
\include{bibtex}

\shorttitle{Growth of Structure from a Cluster Survey}
\shortauthors{Peterson et al.}

\begin{document}

\title{Evidence for Non-Linear Growth of Structure 
from an X-ray Selected Cluster Survey using a
Novel Joint Analysis of the Chandra and XMM-Newton Archives}

\author{J. R. Peterson$^1$, J. G. Jernigan$^2$, R. R. Gupta$^3$, J. Bankert$^1$, S. M. Kahn$^3$}

\affil{$^1$ Department of Physics, Purdue University, 525
    Northwestern Avenue, West Lafayette, IN 47907}

\affil{$^2$ Space Sciences Laboratory, University of California, Berkeley, CA
  94720}

\affil{$^3$ Kavli Institute for Particle Astrophysics and Cosmology,
  Stanford University, PO Box 20450, MS 29, Stanford, CA 94309}

\email{peters11@purdue.edu, jgj@ssl.berkeley.edu, ravgupta@physics.upenn.edu,
  jbankert@purdue.edu, skahn@slac.stanford.edu}

\begin{abstract}
We present a large X-ray selected serendipitous cluster survey based on a novel
joint analysis of archival Chandra and XMM-Newton data.
The survey provides enough depth to reach
clusters of flux of $\approx 10^{-14}~\mbox{ergs}~\mbox{cm}^{-2}~\mbox{s}^{-1}$
near $z$ $\approx$ 1 and simultaneously a large enough
sample to find evidence for the strong evolution of clusters expected from structure
formation theory.  We detected a total of 723 clusters
of which 462 are newly discovered clusters with greater than 6$\sigma$ significance.
In addition, we also detect and measure 261 previously-known clusters
and groups that can be used to calibrate the survey.
The survey exploits a technique which
combines the exquisite Chandra imaging quality with the high throughput of the
XMM-Newton telescopes using overlapping survey regions.   A large fraction of the contamination from AGN point
sources is mitigated by using this technique.  This results in a higher sensitivity for
finding clusters of galaxies with relatively few photons and a large part of
our survey has a flux sensitivity between
$10^{-14}$ and $10^{-15}~\mbox{ergs}~\mbox{cm}^{-2}~\mbox{s}^{-1}$.  The survey covers
41.2 square degrees of overlapping Chandra and XMM-Newton fields and 122.2
square degrees of non-overlapping Chandra data.  We measure the log N-log S
distribution and fit it with a redshift-dependent model characterized by a
luminosity distribution proportional to $e^{-\frac{z}{z_0}}$.  We find that
$z_0$ to be in the range 0.7 to 1.3, indicative of rapid cluster evolution, as
expected for cosmic structure formation using parameters appropriate to the
concordance cosmological model.  Confirmation of our cluster detection
efficiency through optical follow-up studies currently in progress will help
to strengthen this conclusion and eventually allow to use these data to derive
tight contraints on cosmological parameters.
\end{abstract}

\keywords{cosmology:observations --- galaxy:clusters --- X-rays:galaxies:clusters}

\section{Introduction}

There is a long history of compiling catalogs of nearby galaxy clusters by using
optical telescopes \citep{abell,zwicky}.  The development of focusing
X-ray telescopes and the early discovery of X-rays from clusters of galaxies
\citep{byram} provided another sensitive method for finding clusters of galaxies.
A number of X-ray catalogues now exist for the brightest clusters.
Most of the nearby clusters are known from flux-limited surveys that cover a large fraction of
the sky from ROSAT and Einstein down to
$10^{-12}~\mbox{ergs}~\mbox{cm}^{-2}~\mbox{s}^{-1}$ (REFLEX, \citealt{boehringer};
BCS, \citealt{ebeling}; EMSS, \citealt{gioia90}).
These surveys have discovered 100-450 clusters.   Deeper surveys that
extend to $10^{-14}~\mbox{ergs}~\mbox{cm}^{-2}~\mbox{s}^{-1}$ and therefore
probe a larger range of redshifts have found 60-200 clusters serendipitously
(160 square deg., \citealt{vikhlinin}; 400 square deg. \citealt{burenin};
  ROSAT NEP, \citealt{gioia01}, \citealt{henry2};
WARPS, \citealt{scharf}, \citealt{horner}).
XMM-Newton surveys have found 19 clusters serendipitously
\citep{georgantopoulos} and 12 clusters in the Large Scale Structure Survey
\citep{willis} down to deeper flux levels.

The slope of the log N-log S distribution for clusters is close to -1
down to a flux limit of $\approx 10^{-13}~\mbox{ergs}~\mbox{cm}^{-2}~\mbox{s}^{-1}$
which means that a wide survey with short exposures will 
approximately detect the same number of clusters as a narrow survey
with longer exposures for the same total exposure time.
Thus the total number of clusters is simply
proportional to the product of a telescope's etendue (effective area times
field of view) times the total survey exposure time divided by the minimum
number of photons that one can reliably use to identify a cluster
unambiguously.  A large number of the previous surveys used ROSAT which has an
area of 200 cm$^2$ and a field of view of 3 square degrees.
The fields of view of Chandra (0.1 square degrees) and XMM-Newton
(0.2 square degrees) are not particularly high, but the effective area of
Chandra (400 cm$^2$ at 1 keV) and especially XMM-Newton (2000 cm$^2$ at 1 keV)
are high.  XMM-Newton, in particular, has an etendue comparable to ROSAT.  One
difficulty with using XMM-Newton to perform cluster surveys compared to ROSAT
is that the large area and smaller field of view will collect on average more
distant groups of galaxies, with smaller angular size, because the survey will be deeper
rather than wider for the same exposure time.  The fact that the clusters have
smaller angular size means that it is harder to determine the difference between a cluster or
group and a point source.  On the other hand, Chandra has
a exquisitely sharp point-spread function (PSF)
that distiguishes AGN point sources from clusters at all
redshifts across its entire field of view.  Therefore, by implementing an analysis
method which combines the Chandra PSF with the XMM-Newton throughput, we
demonstrate that we can produce a large sensitive X-ray cluster survey.

Recently, a number of independent astronomical methods have converged on a standard
cosmological model, which indicates the presence of a poorly understood dark
energy component.  An extensive search for new galaxy
clusters will provide new constraints on cosmological
models.  In general, the abundance and distribution of clusters of galaxies in the
Universe provides precision and complementary cosmological constraints to other
methods \citep{eke}. In particular, the measurements are most sensitive to $\sigma_8
\Omega_m$, where $\sigma_8$ is the normalization of the matter power spectrum
on 8 Mpc scales and $\Omega_m$ is the matter density.  A large
enough sample at a range of redshifts can constrain $w$, the equation of
state parameter for dark energy (e.g., \citealt{haiman}, \citealt{wang}).
The distribution of clusters as a function of mass and
 the spatial power spectrum constrains cosmological parameters as well.

The content of this initial paper principally concerns the exposition of the new approach for
detecting a large sample of X-ray clusters based on a combined analysis of the all useful
data in the full Chandra and XMM-Newton archives. We have discovered a new unbiased sample
of 462 clusters that reaches the most sensitive flux limit achieved to date
 (~$10^{-14}~\mbox{ergs}~\mbox{cm}^{-2}~\mbox{s}^{-1}$).
We summarize the cosmological implications of this new information with an initial
limited  approach that assumes the concordance model of the universe is correct
and compares the measured Log N-Log S curve of the new X-ray cluster sample with what would
be expected in a concordance Universe. The limited scope of this approach precludes
a full analysis of the consistency of these new data and the concordance model.
We do confirm approximate consistency with the concordance model in two independent ways.
First we show that the rapid deviation of the Log N-Log S curve for fluxes less than
 (~$10^{-13}~\mbox{ergs}~\mbox{cm}^{-2}~\mbox{s}^{-1}$) for a universe without evolution of
the density of clusters is consistent with an expected exponential decrease in cluster density
for clusters near z $\approx$ 1 and beyond. This result approximately matches the expected
density for clusters of mass range expected at a distance of z $\approx$ 1 (see figure 15). 
Secondly, assuming a concordance model, we compute the expected Log N-Log S curve
and find approximate agreement with the newly measured Log N-Log S curve (see figure 16).
The detailed explanation of these two comparisons assuming a concordance model
appears later in this paper. A full analysis beyond approximate consistency with
the concordance model is beyond the scope of this initial paper.
In particular, we must improve estimates of several possible systematic errors including
the volume of the surveyed region in three dimensions,
the exact detection threshold near the flux limit,
the re-calibration of the temperature, size, luminosity and mass relations
of X-ray selected clusters.
We present the log N-log S of the unbiased sample of 462 new clusters without
redshift measurements because redshifts (and thus luminosities)
are not needed for this initial simple analysis.  
Future work will address the full cosmological implications of these measurements
beyond just the Log N-log S curve for clusters.

\section{Basic Methodology}
 
 X-ray observations
that have both XMM-Newton and Chandra data for the same piece of the sky have the advantage that they can use both
the Chandra PSF and the XMM-Newton throughput.  For example, finding three photons
within one or two arcseconds of each other in a 5 ks Chandra exposure can
umambiguously identify an AGN.  The probability of background fluctuation or a
diffuse source creating that situation is negligible.  The number of photons required to
determine a point source in XMM-Newton is much higher.

On the other hand, XMM-Newton has $\sim$5 times the effective
area and twice the field of view of the Chandra telescope.  So over both
missions' lifetimes XMM-Newton has collected 10 times as many photons as
Chandra from clusters.  Therefore, by simply using Chandra data to find the
positions of AGNs, we can subtract photons from the XMM-Newton data and select
clusters with as few as 10 to 15 photons.  We therefore have a sensitivity several
times higher than if we used XMM-Newton data alone.  It may be possible to use
the full XMM-Newton dataset after calibrating
the AGN contribution precisely.  The major
challenge is to learn how to combine and calibrate all the data and correct
for the non-uniformity of the survey.  New analysis techniques are
required, since the exposures are very deep and contain many smaller clusters
at high redshift.

\section{Cluster Selection Procedure}

Following we describe the basic data processing, image reconstruction, point
source removal procedure, and source detection and selection methods.

\subsection{Basic Data Processing}

\begin{figure*}[htb]
  \begin{center}
      \plotone{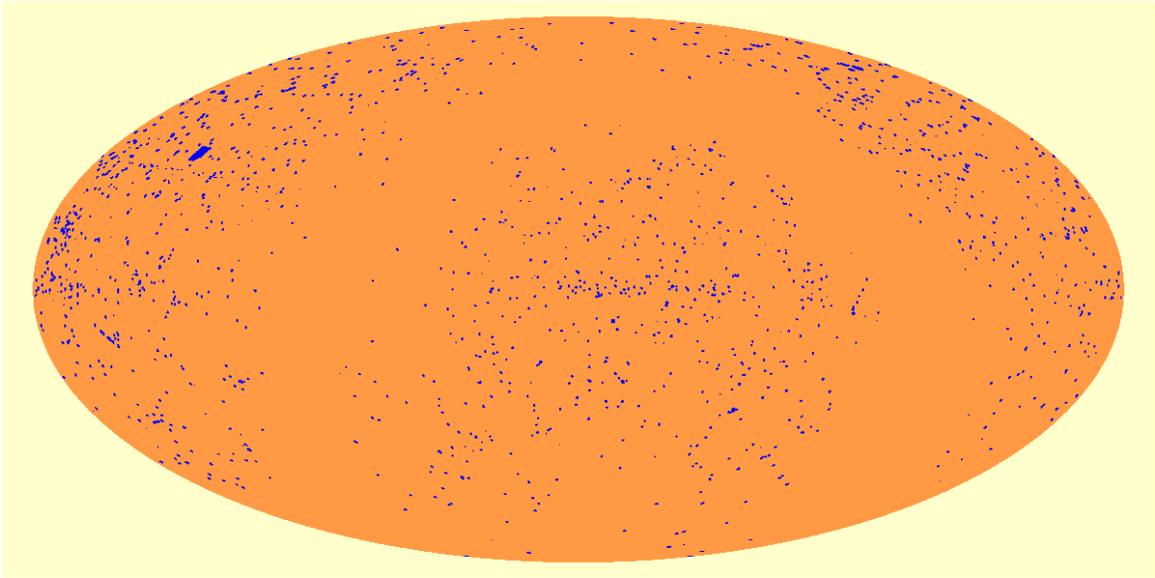} 
  \end{center}
  \caption{\label{fig:survey}  A map of the survey in equatorial coordinates.
  The blue regions mark areas where there is either Chandra data or
  overlapping Chandra and XMM-Newton data.  Approximately 0.4\% of the sky is
  covered by this survey.  Notably absent is the galactic plane data.
  The largest contiguous patch in the upper left is the Bootes Chandra survey.
  }
\end{figure*}

To start this project, we first scanned the XMM-Newton archive for
any fields that could potentially overlap with Chandra fields for all data in
either archive as of September 2006.  Using these fields
we can combine the advantages of the Chandra PSF and the XMM-Newton throughput
in the following method.  A remarkably
large fraction (nearly half) of the observations overlap with Chandra
observations in at least some part of their field of view.  This project clearly
benefits from the fact that X-ray observers tend to choose similar targets
with both observatories.
We downloaded 1201 XMM-Newton fields that overlapped with the
the 2706 Chandra ACIS-S or ACIS-I fields.  We excluded fields within 10 degrees of the Galactic plane
or within 5 degrees of the SMC and LMC.   We also included fields that have only Chandra data, since
although they will not benefit from the XMM-Newton throughput they can be used
to augment the survey with more data with high PSF quality.  The positions of
the observations are shown in Figure 1.  The survey contains approximately
122.2 square deg. of Chandra-alone data with an average exposure of 44 ks and
41.2 square deg. of overlapping XMM-Newton and Chandra data with an equivalent
Chandra exposure of 166 ks.  Thus, the Chandra-alone data contains about
44\% of the cluster candidates and the overlapping survey will contain about
56\% because of the relative effective etendues of each sub-survey.  A total
of 54 Ms of Chandra data and 7 Ms of XMM-Newton data were used.

We used the Chandra level 2 events
and applied the standard destreak processing tool.  We processed all the
XMM-Newton data using the SAS pipeline.  The Chandra photon events were selected to
have corrected energy (PI) between 0.5 and 7 keV, have no bad data quality flag,
and have event pattern less than or equal to 4.  The XMM-Newton photon events
were selected to have corrected energy (PI) between 0.5 and 7 keV and an event
grade of 0, 3, or 4.
Background flares were removed from all the 3907 datasets by first binning the events
in 100 second time bins.  We then iteratively removed 100 second intervals
which have the highest count rate until the signal squared (proportional to
the number of time bins) divided by the noise (proportional to the total
counts) was maximized.  This procedure optimally removed the variable component of the
background (or a bright flaring AGN).  We also removed observations with
effective exposures (in ks) divided by the count rate greater than 0.2.  This
removed extremely short exposures that almost entirely contain background flares.

\subsection{Image Reconstruction}

We scanned the entire sky and identified all photons from any of the datasets
that could have originated in a given grid cell of size 1 square degree.  On
average there are usually about 2 different observations
from either telescope that overlap with a given field position.  Within each 1 square
degree we only include cluster candidates in the catalog if they fall within
the central 0.5 degree by 0.5 degree box, but we analyzed the entire set of
photons in the box through the procedure we describe below.  The 1 degree
square regions were chosen, however, so that they overlapped by 0.5 degrees,
leaving no residual binning artifacts.

We only included events from the Chandra data that were within 15 arcminutes of
the center of the field of view.   This was necessary because the size of the PSF increases
significantly at larger angles, but still a large fraction of the data were
included.  This angle cut also means that some ACIS-S data was included with
ACIS-I pointed observations and some ACIS-I data were included with ACIS-S
pointed observations.  For some particularly deep observations with long
exposures, we only included a randomly chosen set of $10^7$ photons and then
scaled the measurements accordingly, since some of the algorithms we employ
below have run times proportional to the number of photons squared.  This was
rare and only occurs for less than $1\%$ of the datasets.

For every arcminute of the sky, we computed the effective exposure times
telescope area (having units $\mbox{cm}^2~\mbox{s}$) including the effects of telescope vignetting.  To do this, we
assume that the ratio of effective areas of the ACIS-S, EPIC-PN, and EPIC-MOS
were, respectively,  1.4, 4.8, and 1.6 times the ACIS-I area and then we scaled all results of the effective
area exposure map to ACIS-I.  In reality, the relative area depends on the
area of the source, but the values we found were an empirically best averages
over all sources and background.  Similarly, for the vignetting profile, we
assumed a Lorentzian profile of the form $1/(1+(\theta/\theta_0)^2)$ where $\theta_0$ is
15 arcminutes for XMM-Newton and 25 arcminutes for Chandra.  This roughly matched the
average energy dependence of the vignetting.

We then removed XMM-Newton photons and set the XMM-Newton exposure-area to zero
where there were no Chandra data in either that location in the 1 arcminute exposure-area map or in the adjacent 4
squares in the 1 arcminute exposure-area map.  This then conservatively
removed any edge effects, and we are left with XMM-Newton events only where
there is a complete set of Chandra events.  A sample from the Lynx
field is shown in Figure 2 after this step where there is a deep exposure with
both Chandra and XMM-Newton.  After this calculation, we have an estimate of
the exposure-area for every square arcminute of the entire sky.  For each
arcminute of the sky, we also computed the distance to the aimpoint of any
exposure for both Chandra and XMM-Newton exposures, in order to compare our
candidate sources with an estimate of the PSF size.  We also estimated the
average count rate for each square degree of the sky for background estimation.

\subsection{Point Source Removal}

\begin{figure*}[htb]
  \begin{center}
    \plottwo{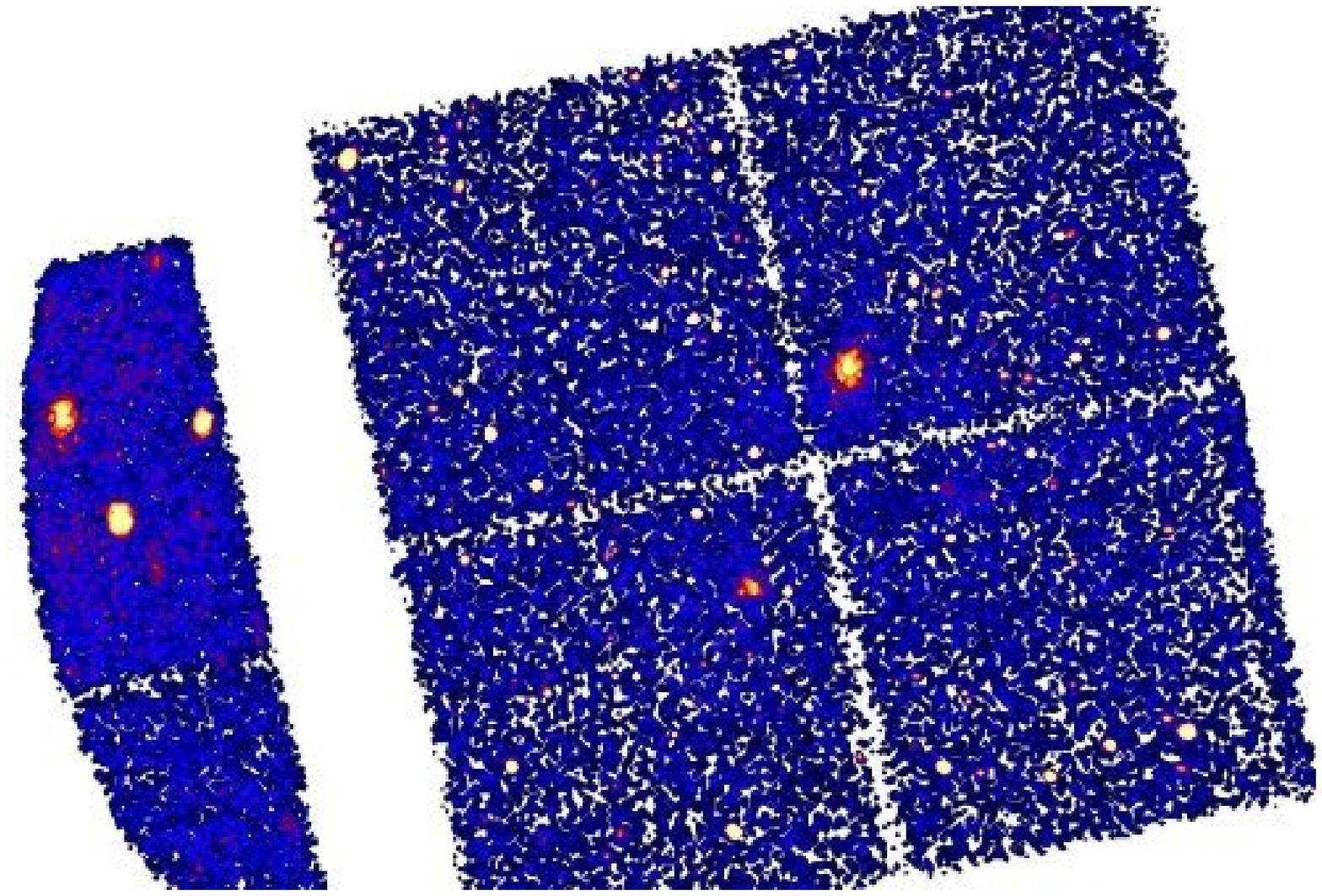}{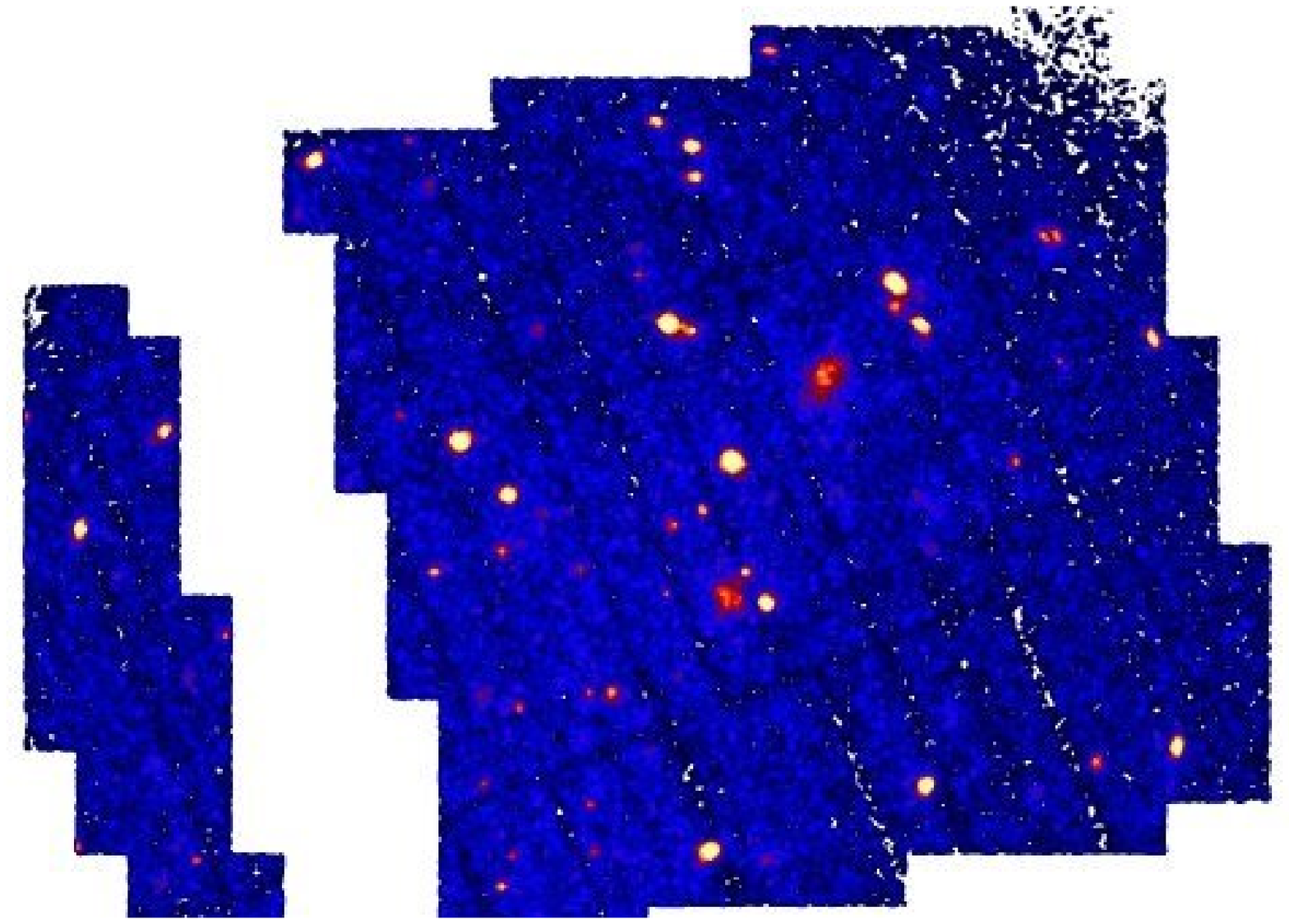} 
  \end{center}
  \caption{\label{fig:image1}  A Chandra (left) and XMM-Newton (right)
  exposure of the Lynx field after the non-overlapping XMM-Newton data has
  been removed.  The Chandra field is mostly an ACIS-I pointing and a small
  piece of the ACIS-S array on the left.  This is an example of a deep
  exposure in the survey with both XMM-Newton and Chandra data. This pixel
  scale is 5 arcseconds, and the brightest regions correspond to roughly 100
  counts per pixel.}
\end{figure*}

\begin{figure*}[htb]
  \begin{center}
      \plottwo{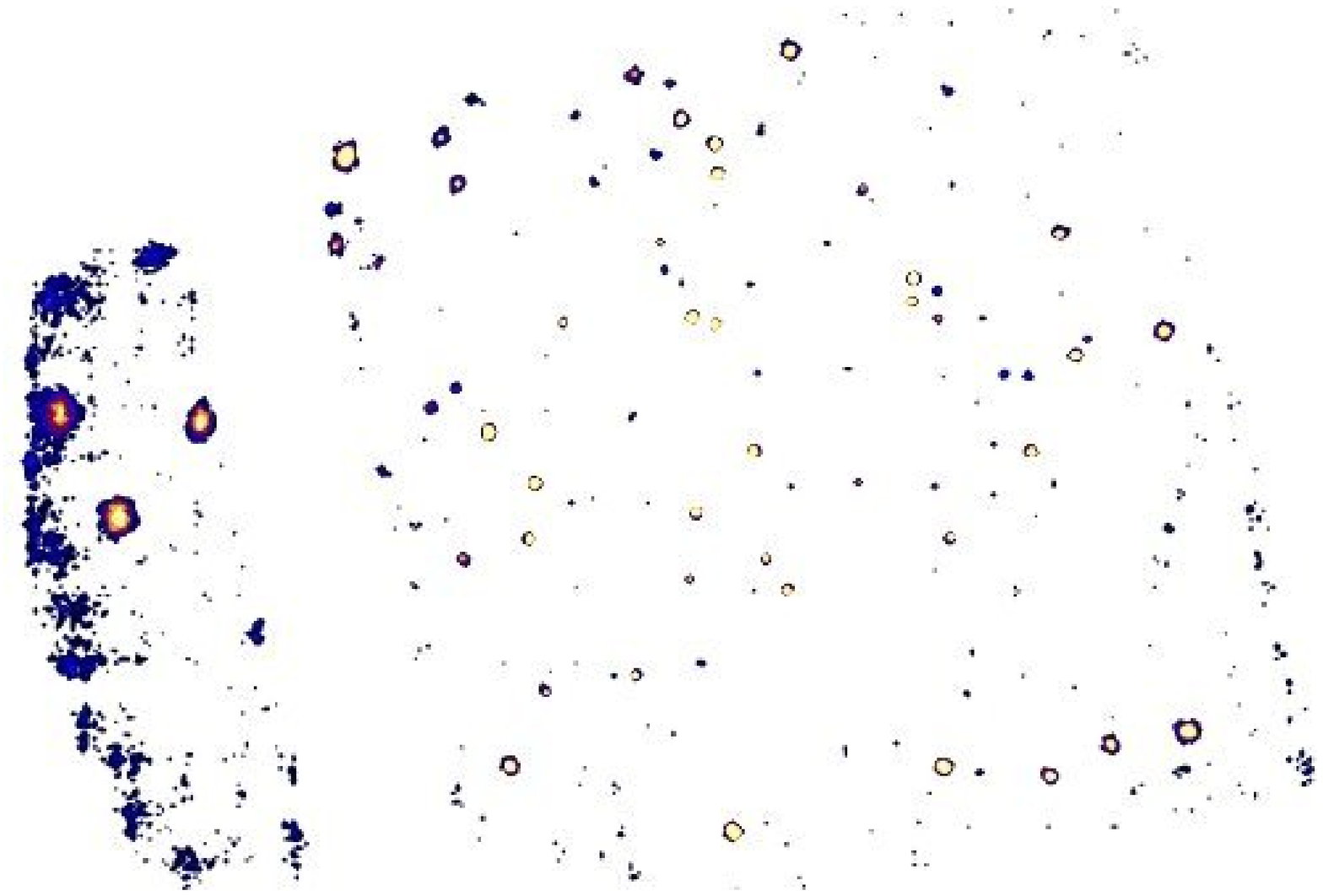}{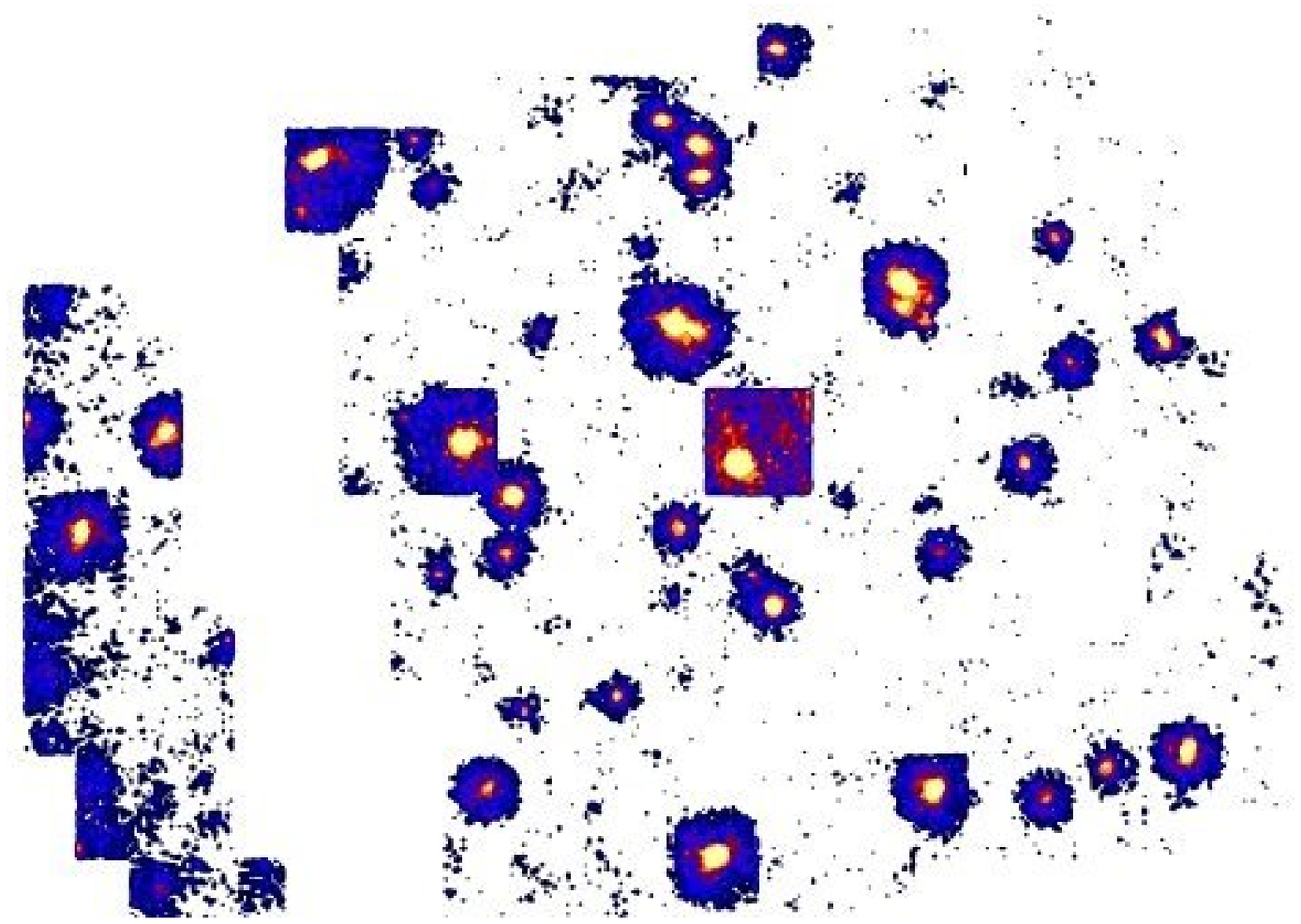}
      \plottwo{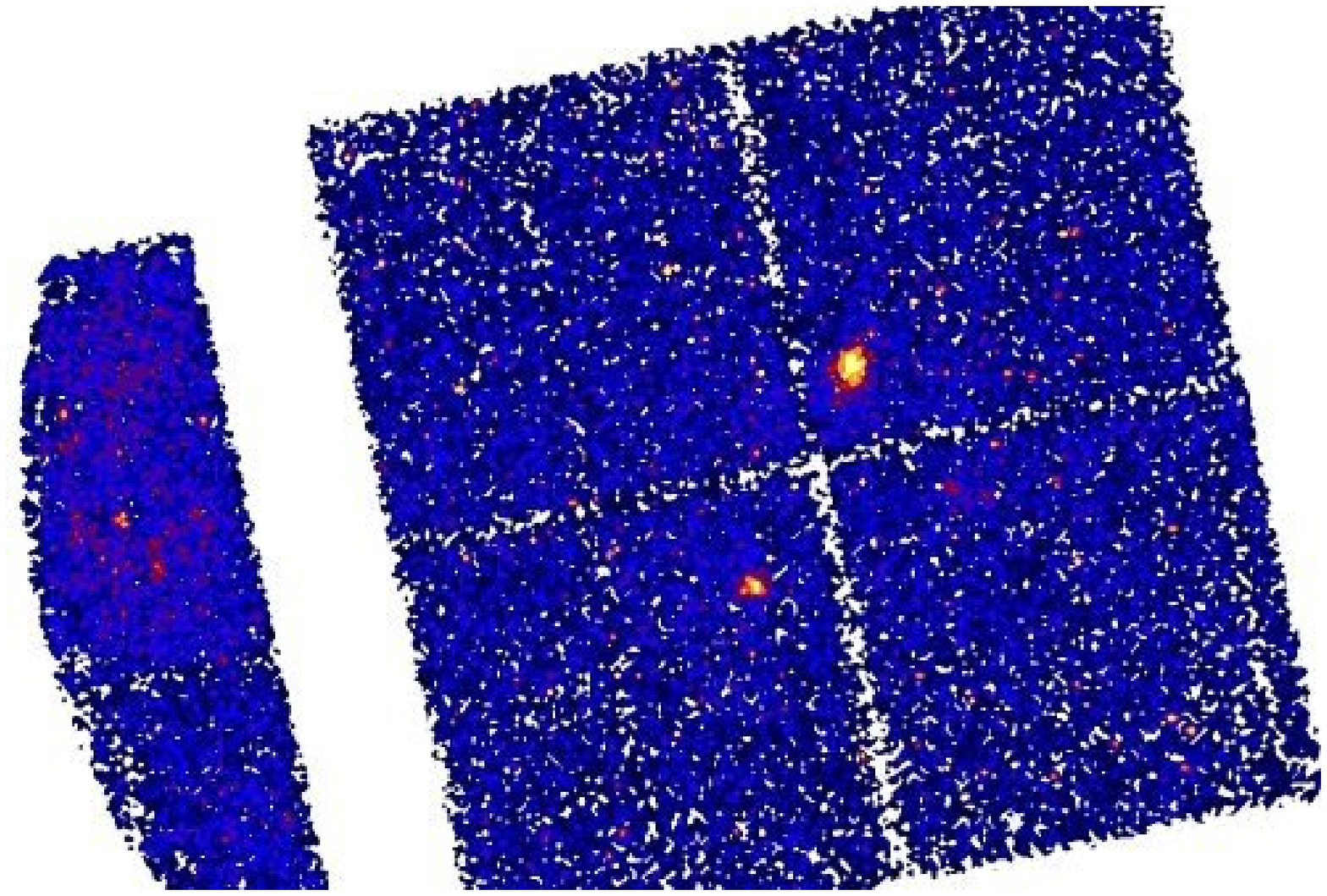}{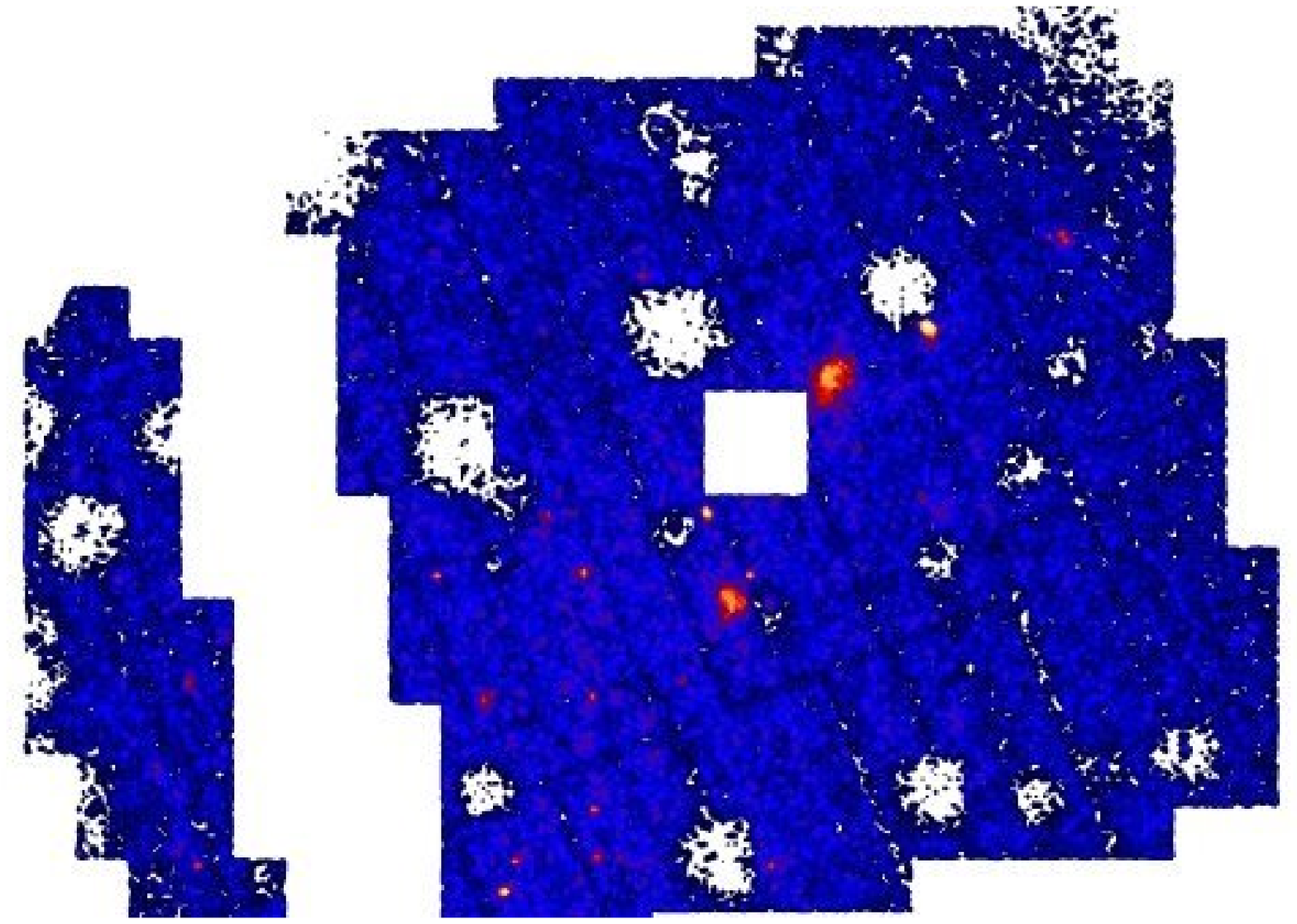} 
  \end{center}
  \caption{\label{fig:image2}  The upper left panel shows the AGN photons
  removed from the Chandra data using the method described in the text.  The
  lower left panel contains the cluster candidate emission and background
  events.  The upper right panel shows the corresponding XMM-Newton photons
  that were subtracted because they were at close to the same position as the
  Chandra data.  The lower right panel shows the XMM-Newton photons after
  subtracted.  The subtraction is not perfect, since the AGN sources vary and
  the method may over or undersubtract some AGN.
  However, it is clear most of the AGN emission complicating the source
  finding in the right in Figure 2 has been removed.}
\end{figure*}

\begin{figure*}[!htb]
  \begin{center}
      \plottwo{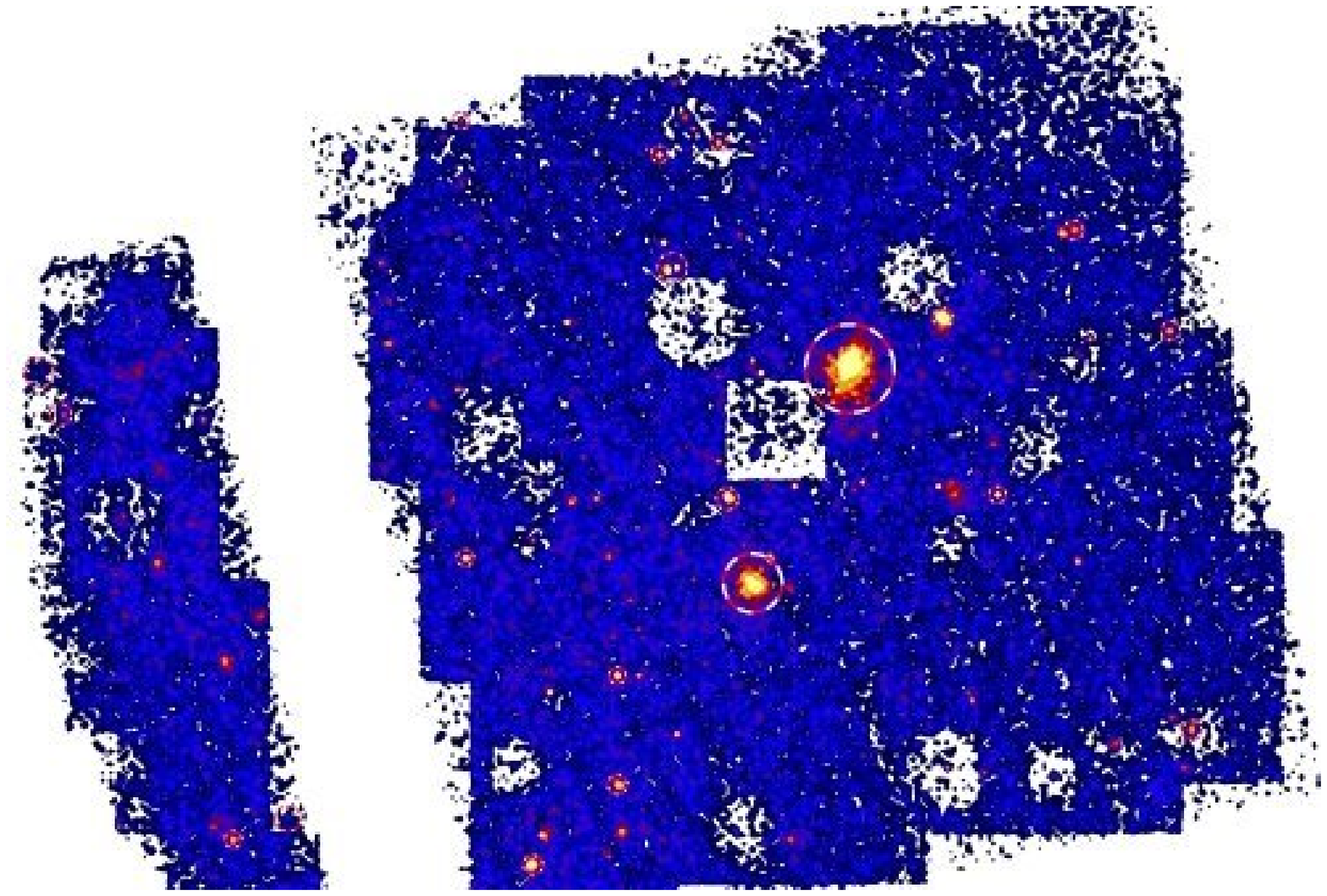}{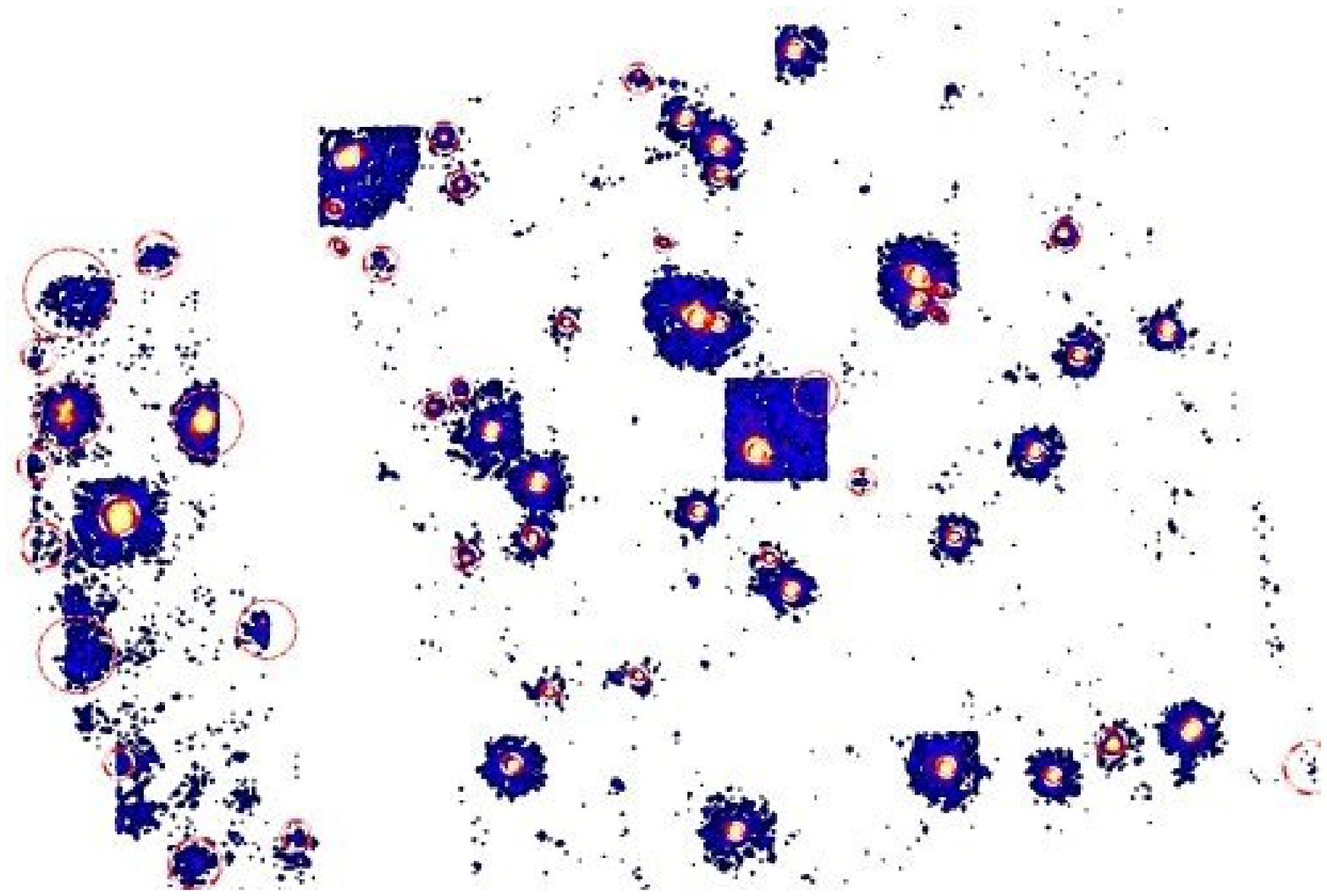} 
  \end{center}
  \caption{\label{fig:image4}  The cluster candidate and background photons
  obtained by adding the left images above is shown on the left.  Similarly,
  the co-added AGN emission is shown on the right.  The cluster and AGN candidates are
  then found on these maps using the wavelet method described in the text.
  The candidates are circled with a size proportional to the size determined
  from the wavelet methods.  The thicker circles correspond to those that
  satisfy all remaining data cuts.  In this case, the two prominent clusters
  near the center are selected.}
\end{figure*}

\begin{figure*}[!htb]
\begin{center}
\plottwo{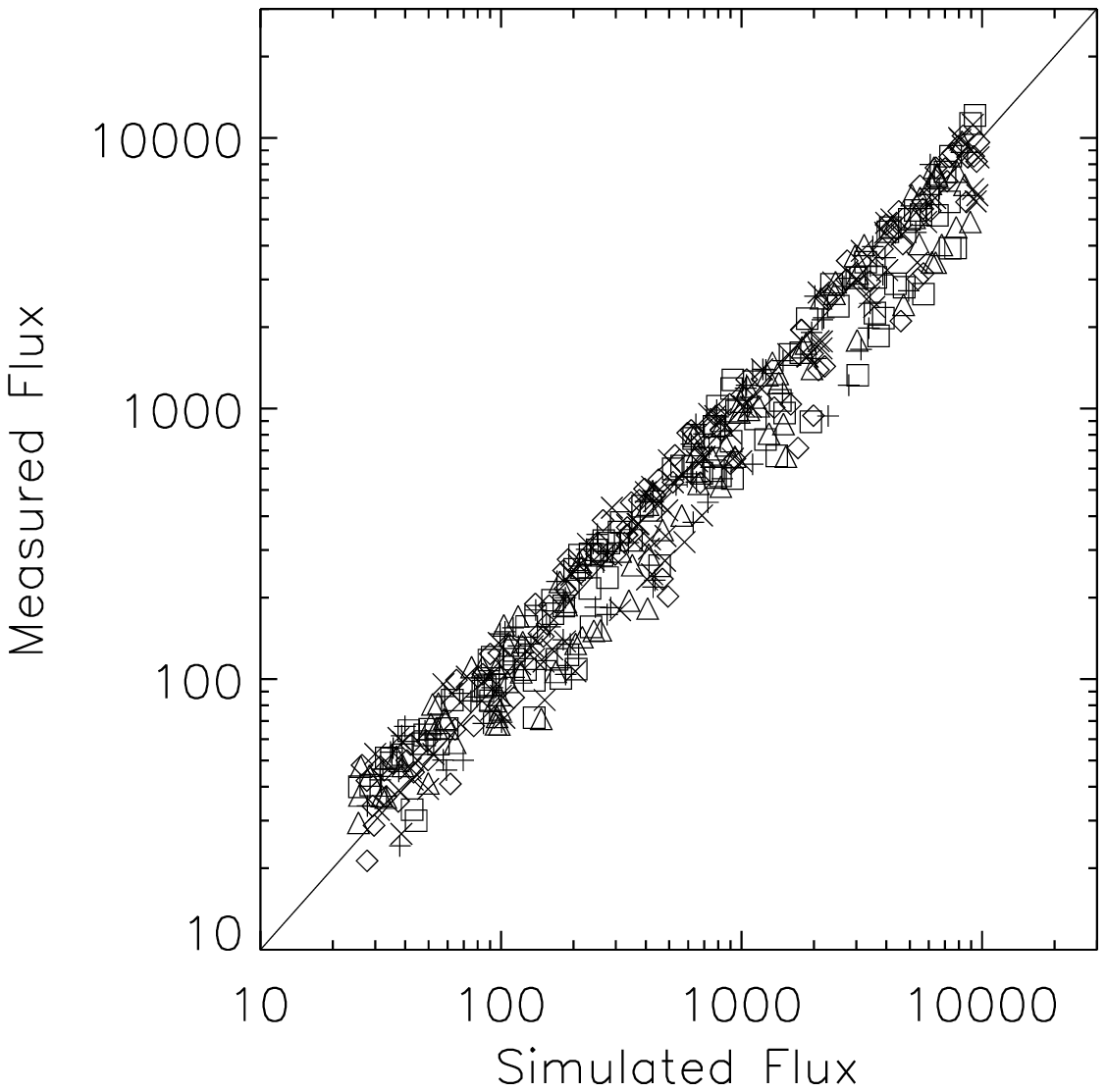}{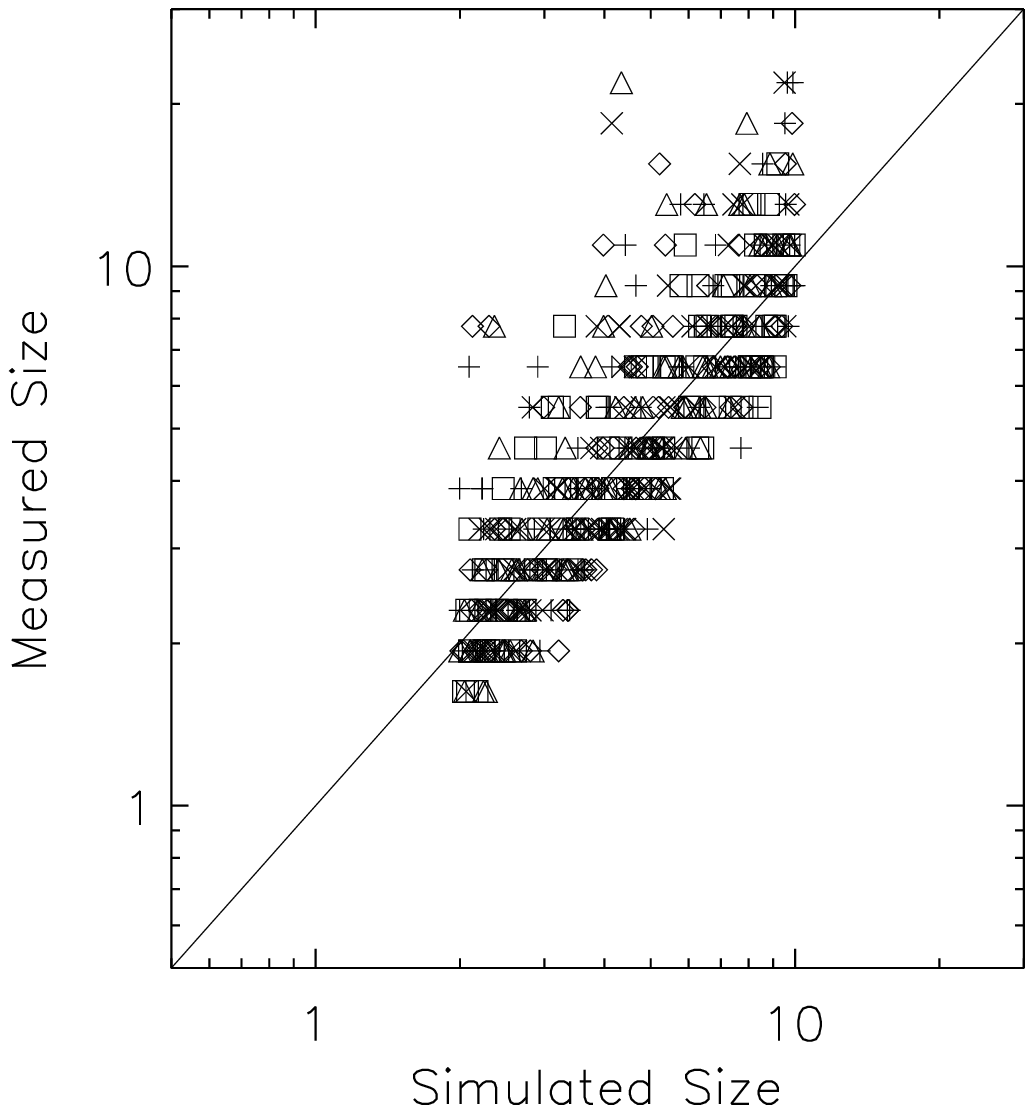}
\end{center}
\caption{\label{fig:fluxcalib}  The source flux of the simulated clusters
  compared to the measured flux using the source detection algorithm as
  described in the text.  Each point represents a different simulated cluster
  with a different background level (diamonds=$10^{-6}$, squares=$10^{-4}$,
  triangles=$10^{-2}$, x sign=$10^{0}$, and plus sign=$10^{2}$ counts per
  pixel).  The source flux is the number of photons we put in image, where as
  the measured flux is the source counts estimated by the wavelet method.  The
  source size is the core radius of the beta model and the measured size is
  the most significant size of the wavelet method times the correction factor.}
\end{figure*}

Using the Chandra photons alone we located the point sources by finding photons
where the number of neighbors within the local PSF size exceeded the
number of photons within twice the local PSF size by two sigma.  We only
searched through a subset of photons in the relevant 1 by 1 arcminute block or the
adjacent 1 by 1 arcminute block.  This improved the computation efficiency of
the algorithm.  The local PSF size was estimated to be $(3+20(\frac{\theta}{\theta_1})^2)$
arcseconds where $\theta_1$ is 15 arcminutes.  Photons which satisfied this condition
held were tagged as candidate ``AGN'' photons.  We also randomly untagged a
fraction of these photons in proportion to the relative count rate of photons with
a local PSF size to twice a local PSF size.  This allowed us to fill back in
some photons which may be due to background.  All Chandra photons were then
tagged as either ``AGN'' photons like those in the top left panel in Figure 3
or ``Cluster+Background'' photons like those in the lower left panel in Figure 3.

Then for each Chandra photon we removed a set of the nearest XMM-Newton
photons (when there are any) given by the ratio of the
exposure-area maps we computed earlier.  The nearest XMM-Newton photon was found by
computing, 

$$\sqrt{\frac{\Delta x^2 + \Delta y^2}{(10'')^2} +\frac{\Delta e^2}{(150 \mbox{eV} \sqrt{e_{keV}})^2}}$$. 

\noindent
where $\Delta x$ and $\Delta y$ are the difference in spatial coordinates,
$\Delta e$ is the difference in measured photon energy (in eV), and $e_{keV}$
is the photon energy.  Thus, the selected photons were chosen to have a similar position and have similar energy.  If the ratio of
exposure-area maps was not an integer, we chose random numbers appropriately
to decide whether to add extra photons to obtain the proper normalization.  We also did not use XMM-Newton photons
outside of a 2 by 2 arcminute block as with the previous step for
computational efficiency.  This method then selected the XMM-Newton AGN photons even
if the XMM-Newton PSF was large and if it was difficult to determine if the
source was a cluster or AGN from the XMM-Newton data alone.  We then have two
sets of photons for the XMM-Newton data as well:  those that are removed are
composed primarily of AGN photons (top right in Figure 3), and those that have
not been removed that are either from clusters or background (bottom right in
Figure 3).  The method is imperfect when the AGN flux varies significantly
between observations and due to noise, but we can estimate this by comparing
the two maps.  If we subtracted a lot of ``AGN'' photons at a particular
location and we found a source in the subtracted map then this was likely due
to a subtraction problem.  This is described in more detail in the next section.

\subsection{Source Detection and Selection}

\begin{figure*}[!htb]
  \begin{center}
      \plotone{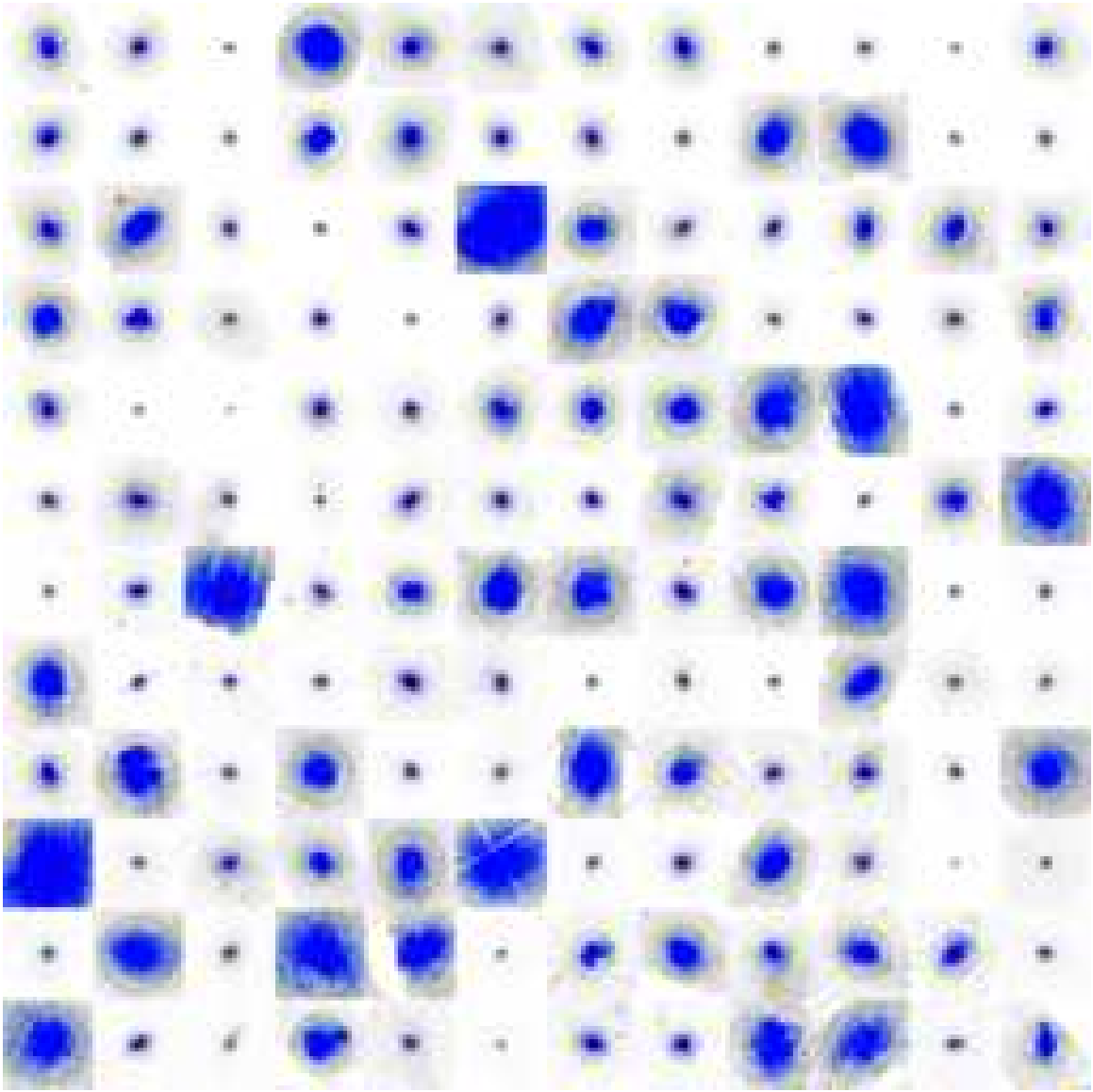}  
  \end{center}
  \caption{\label{fig:gallery1}  The full catalog of cluster candidates
  (continued on next pages).  The
  subtracted AGN emission is shown in red, whereas the cluster emission is
  shown in blue.  The dotted circle represents twice the fitted wavelet size
  of the source detection algorithm.}
\end{figure*}

\begin{figure*}[!htb]
  \begin{center}
      \plottwo{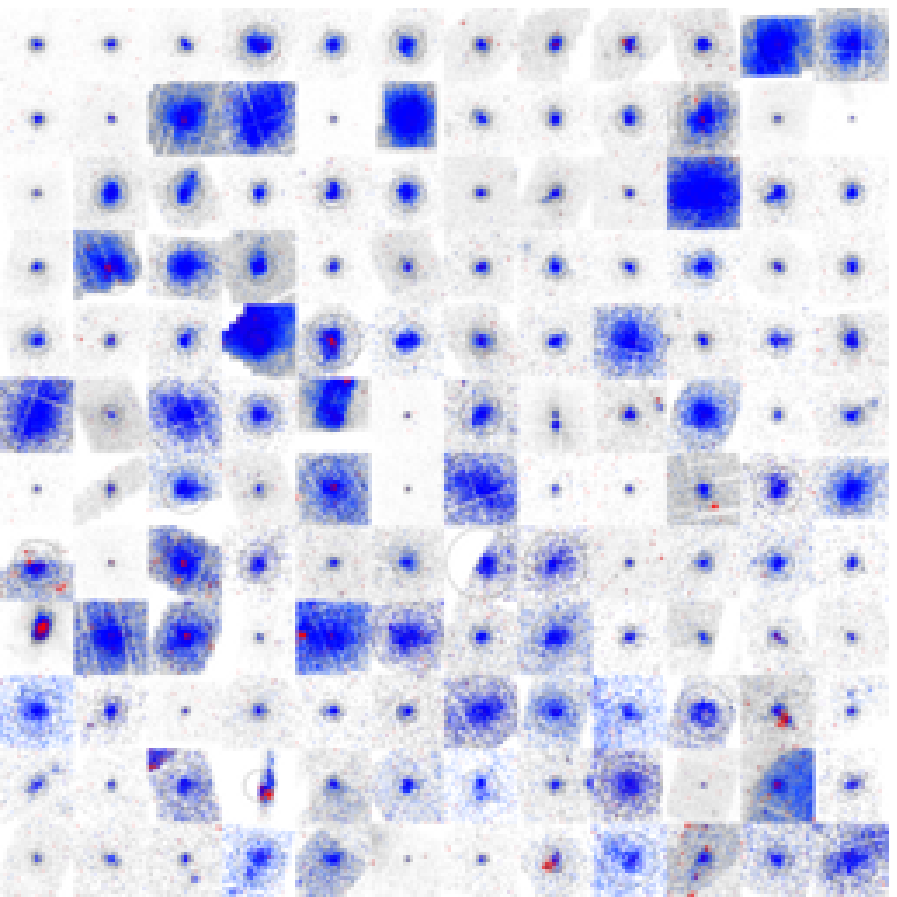}{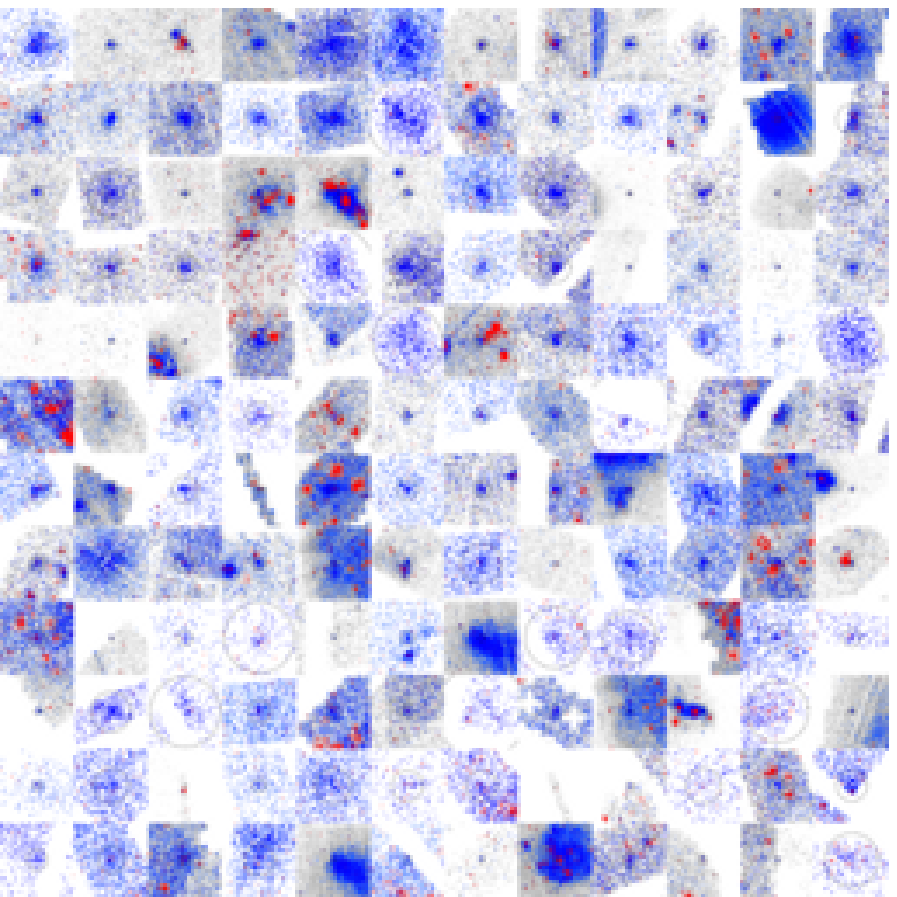}      
  \end{center}
  \caption{\label{fig:gallery2}  The full catalog of cluster candidates (continued).  The
  subtracted AGN emission is shown in red, whereas the cluster emission is
  shown in blue.}
\end{figure*}

\begin{figure*}[!htb]
  \begin{center}
      \plottwo{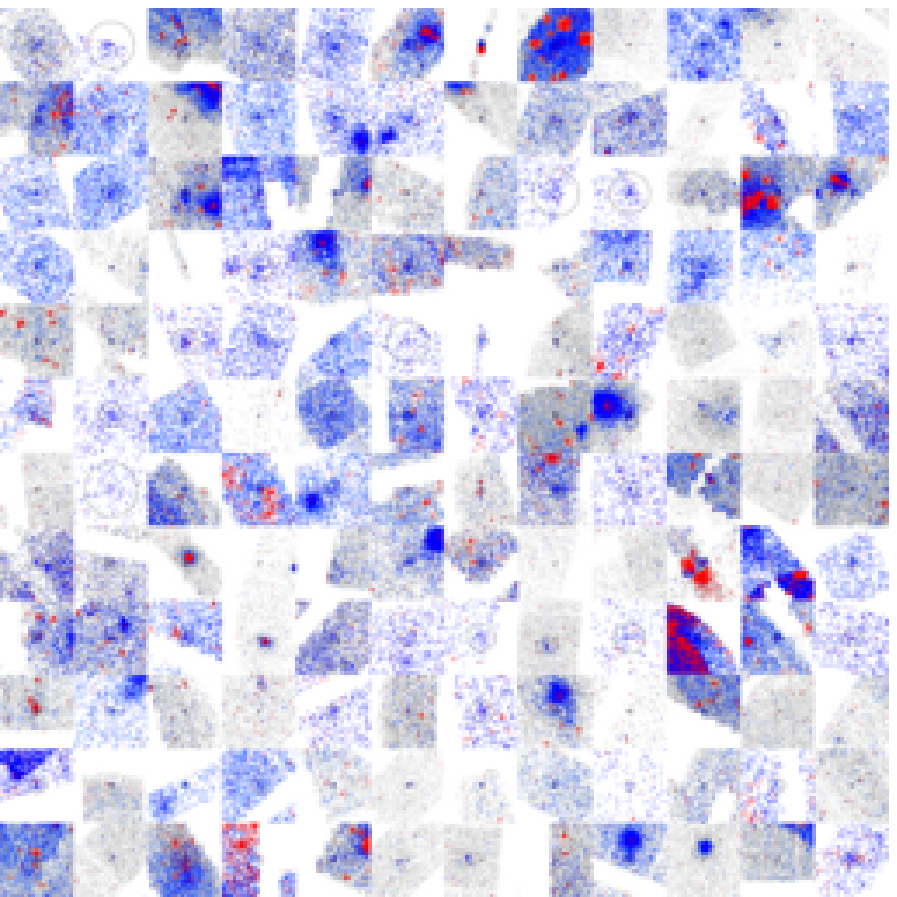}{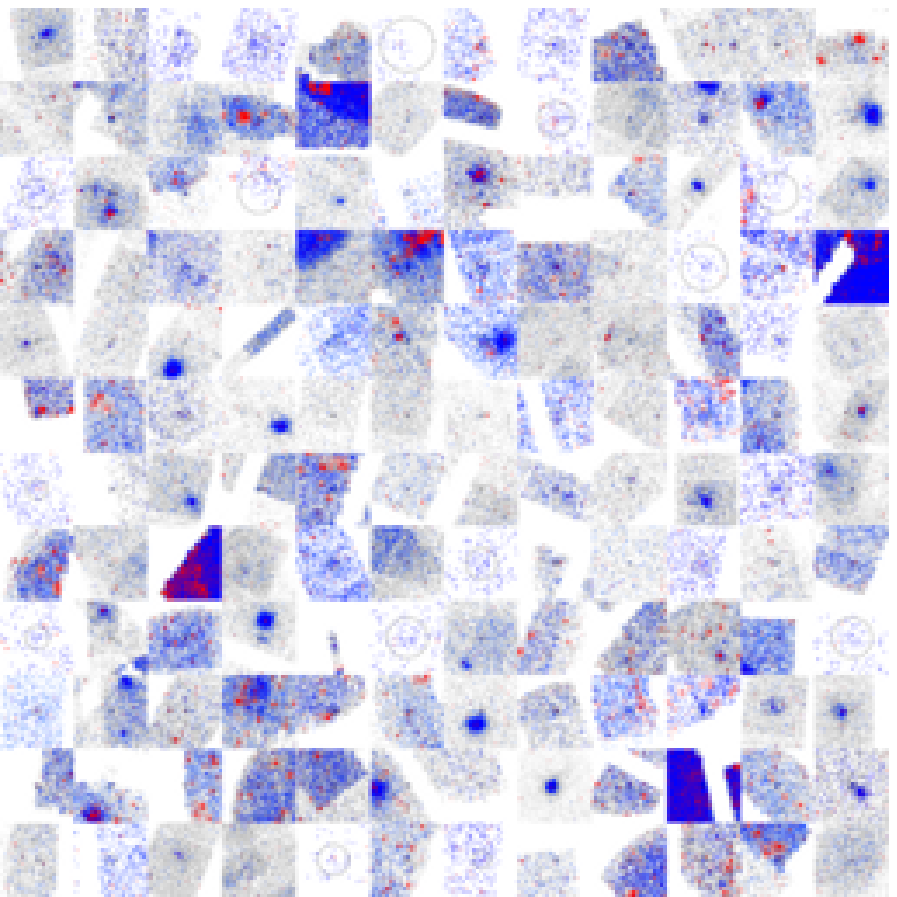}      
  \end{center}
  \caption{\label{fig:gallery3}  The full catalog of cluster candidates (continued).  The
  subtracted AGN emission is shown in red, whereas the cluster emission is
  shown in blue.}
\end{figure*}

\begin{figure*}[!htb]
  \begin{center}
      \plottwo{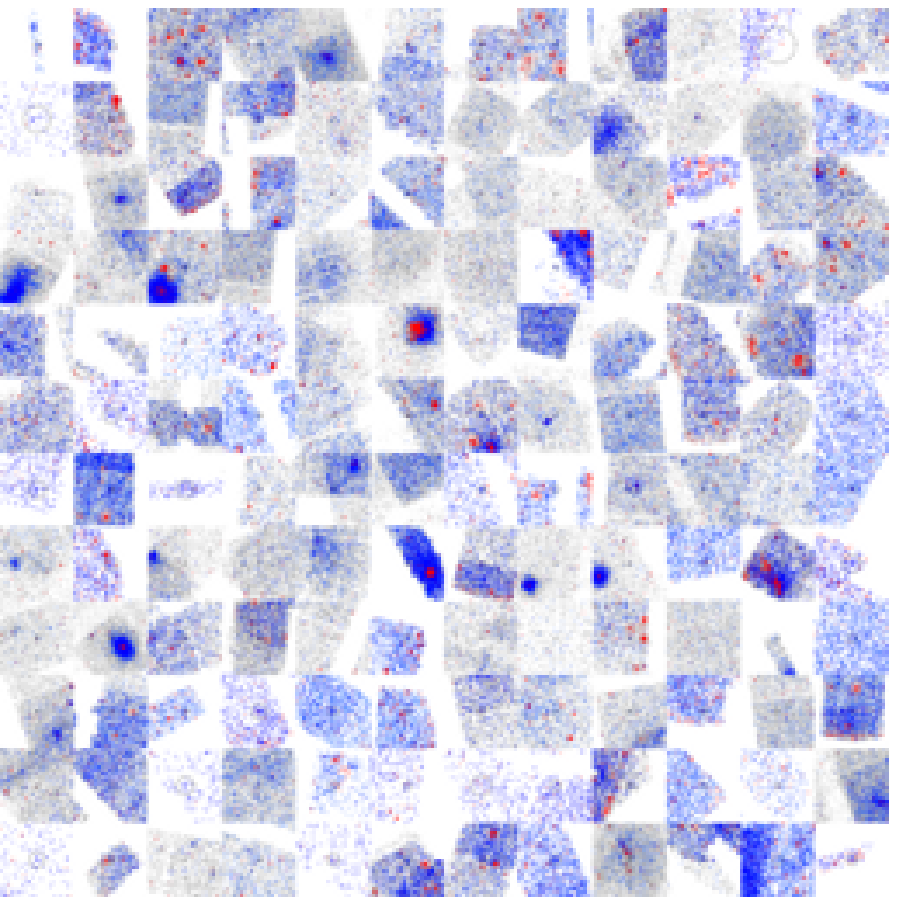}{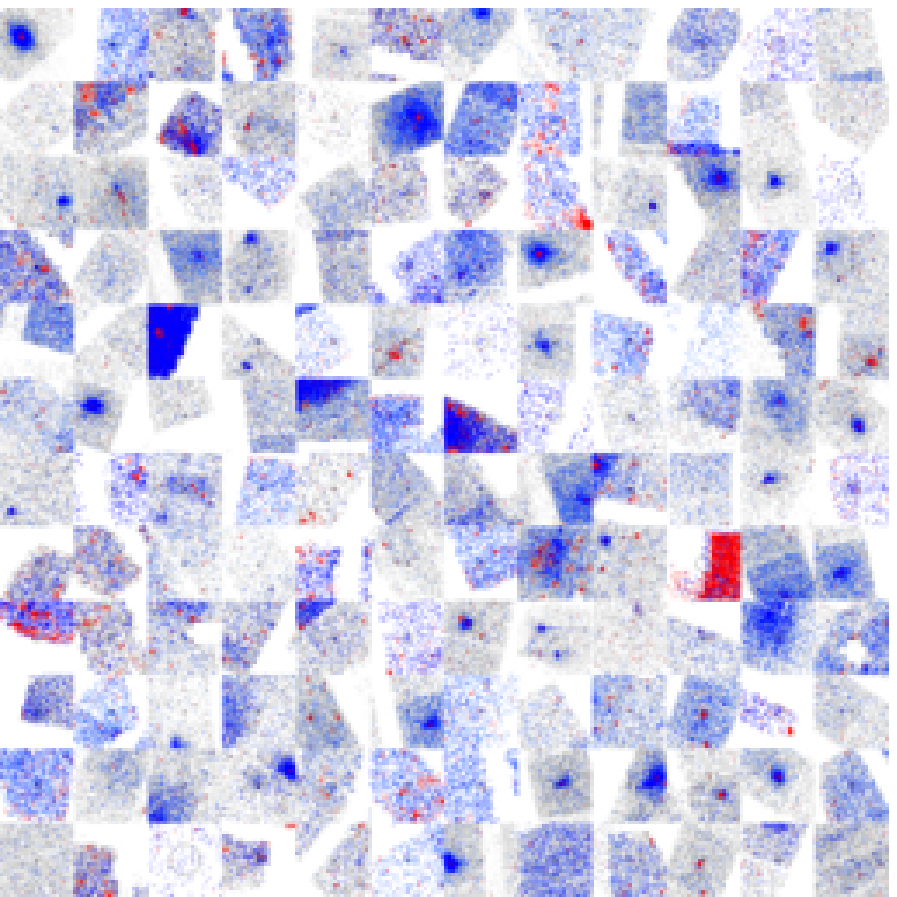}      
  \end{center}
  \caption{\label{fig:gallery4}  The full catalog of cluster candidates (continued).  The
  subtracted AGN emission is shown in red, whereas the cluster emission is
  shown in blue.}
\end{figure*}

\begin{figure*}[!htb]
  \begin{center}
      \plottwo{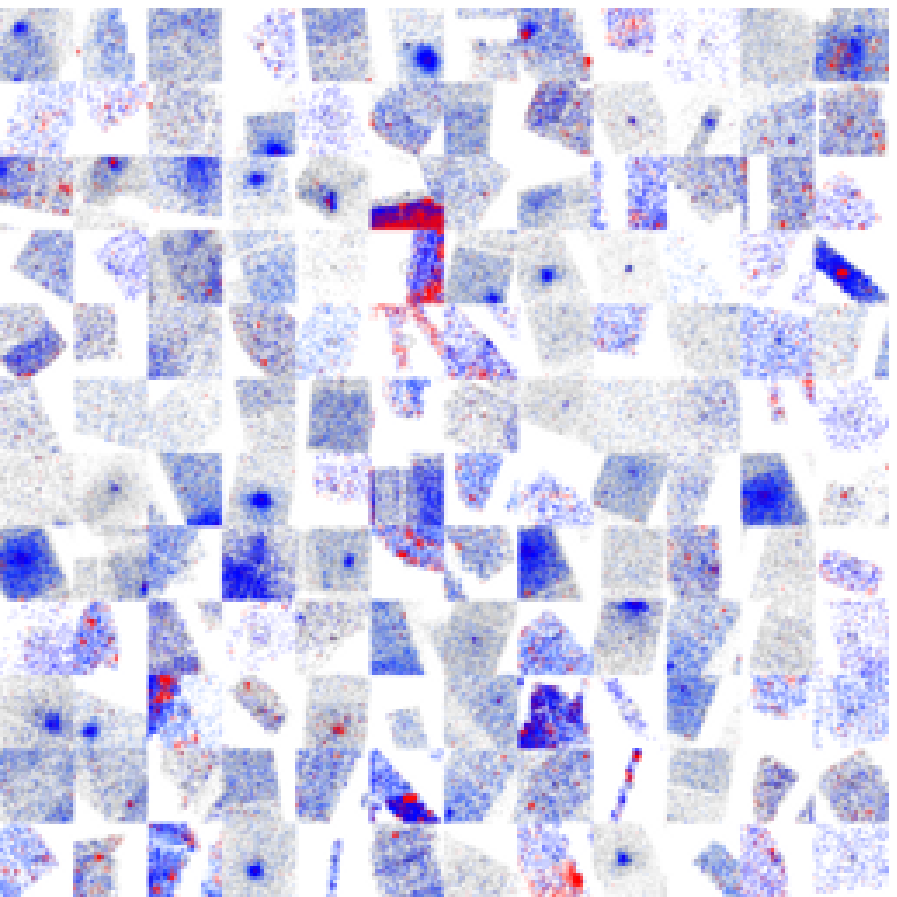}{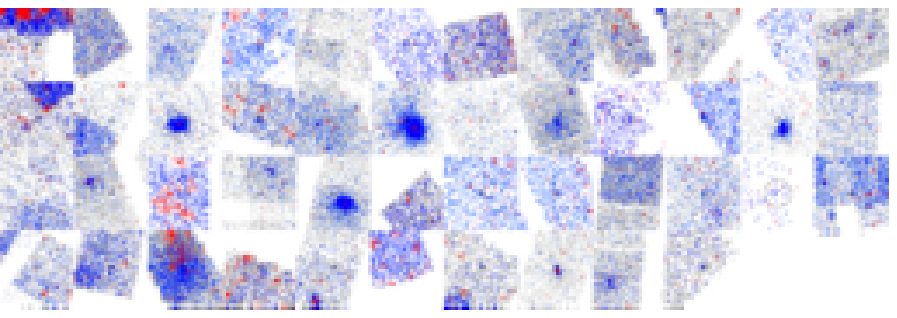}
  \end{center}
  \caption{\label{fig:gallery5}  The full catalog of cluster candidates (continued).  The
  subtracted AGN emission is shown in red, whereas the cluster emission is
  shown in blue.}
\end{figure*}

Next, we constructed the combined maps by adding the XMM-Newton and Chandra maps
for AGN photons and clusters and background photons, respectively as shown in
Figure 4.  We only included photons from 0.5 to 2 keV, since now we wanted to
maximize our sensitivity to finding new clusters and most of the cluster
emission occurs in the soft band.  We binned the photons into a 5 arcsecond grid and
 then find sources using a wavelet method
which is a modified form of that described by Vikhlinin 1998.  To do this, we
computed functions of the following form:

$$ g(x_0, y_0,\sigma) = \frac{1}{2 \pi \sigma^2} \int f(x,y) e^{-\frac{(x-x_0)^2+(y-y_0)^2}{2 \sigma^2}}$$

\noindent
where $g$ is a gaussian convolution of the map $f$ with scale $\sigma$.  These
convolutions can be calculated most efficiently in Fourier space.  An estimate
of the number of photons of a given source at position $(x_0,y_0)$ of size $\sigma$ is given by
 
$$ c(x,y) = \left(\frac{g_{sc}(x_0,y_0,\sigma)}{g_{ea}(x_0,y_0,\sigma)} - \frac{g_{sc}(x_0,y_0,2 \sigma)}{g_{ea}(x_0,y_0,2 \sigma)} \right) $$

\noindent
where the $g_{sc}$ is the integral above using the soft band counts and
$g_{ea}$ is the integral using the exposure-area function.
When the area-exposure map is constant this function will be equivalent to the
normal spherical top-hat function used in wavelet applications.  We stepped
through every pixel of the map using source sizes from 1 to 32 pixels in step
size of $log_2(\mbox{radius})$ of 0.2.  We also estimate the hard band counts (2.0 to 7.0
keV) and the AGN soft band counts at the same position.  The background rate
was estimated locally by convolving a gaussian with $\sigma$ of 480 arcseconds
(3 times the largest source we allow).  We found that the detection significance of a source
could then be estimated by the formula

$$\mbox{significance}  = \frac{c}{n}/\sqrt{\frac{b}{n}} -
4 (b \sigma^2)^{-0.4}$$.

\noindent
where $c$ is the counts function defined in the previous equation, $b$ is the
background function using the previous equation with width of 480 arcseconds, and $n$ is a
normalizing factor $\frac{9}{80 \pi \sigma^2}$.  We
determined this function by performing Monte Carlo simulations of pure poisson
noise.  The first term dominates for high count rates and follows the gaussian
form, and the second term corrects for the fact that at low count rates the
distribution is non-gaussian.  The detection significance, therefore, depends
both on the number of photons in the source and the source size.  We also
found it necessary to calculate the signal to noise as well as the source
significance (signal divided by the square root of noise), because there were highly significant large sources that had low
signal to noise.  We then have a set of candidate sources at various positions
and sizes.  We only select sources with significances greater than 6
sigma and signal to noise greater than one.  These limits were chosen so that we at most
have one spurious source in the entire survey due to background fluctuation and do not include low
signal to noise sources that may be due to non-statistical background
calibration errors.

At this point, the candidate sources could overlap and be at adjacent
locations because they all have enough detection significance.  To select a
unique set of sources, we simply remove any candidate source that is within 3
source radii from another and has lower significance.  The source radius is defined as the
maximum $\sigma$ of the two sources being compared.  We found that this visually
results in only one object being detected per actual source quite efficiently.

To test the ability to correctly measure the flux and size of sources, we
performed a series of photon Monte Carlo simulations.   We first randomly
generated between 30 and 10,000 photons representing a cluster using a
$\beta$ model.  The core radius was randomly chosen between 2 and 10 pixels
and $\beta$ was chosen between 0.5 and 0.75.  We then chose a background level
randomly between $10^{-6}$ and $10^2$ counts per pixel having Poisson noise.
This simulates the wide range of sources and conditions that we might find.  We then used the
source finding algorithm on the simulated image and found that the source flux
was accurately recovered and the source size was accurately measured even in
the presence of the different background conditions.  This is shown in
Figure~\ref{fig:fluxcalib}.  The typical scatter from the true values was
$20\%$ and $30\%$ for the source size and flux, respectively.  We also note
that we recovered 495 out of the 500 sources we simulated.  This indicates
that our source finding algorithm is fairly robust to a range of background
levels and source fluxes. 

We then have a large number of cluster candidates.  We found it necessary to employ four
further selection criteria to remove non-clusters.  First, we computed the
ratio of the counts in the cluster candidate to the counts in the AGN map.  We
required this ratio to be greater than 4.  Without this requirement, AGN which
have increased flux between the Chandra and XMM-Newton exposures will be
undersubtracted.  This is rare since the variability is typically only about
10$\%$, but there are also many AGN candidates.  

We also found it necessary to require that the
source size plus three times the measurement error is at least 4 arcseconds
larger than the local PSF size,

$$\sqrt{(\mbox{size}+3 \sigma_{\mbox{size}})^2-(\mbox{PSF})^2} \geq 4''$$

\noindent
This is a very conservative cut that removes about 30$\%$ of the potential
candidates.  Since, we have added three times the source size error, we are
only effectively removing candidates that have very little chance of actually
being legitimate clusters.  We then expect that only a small fraction of this
30$\%$ would be real clusters or groups that are removed accidentally. The local PSF size is
estimated by combining quadratically the estimate for the local Chandra PSF given in the
previous section with the XMM-Newton PSF in proportion to their exposure area.  The
XMM-Newton PSF is assumed to be 10 arcseconds across the field of view, since it
changes shape but its size does not change dramatically.  The size error is
determined by simply dividing the size by the square root of the number of counts.  Thus, by using the AGN photon subtraction
method above we have removed the vast majority of AGN emission but there is
some residual background due to AGN variability.  Although this limits
our ultimate sensitivity, our method is several times more sensitive than just
finding sources and estimating the size without the photon subtraction.

We also estimate the ellipticity of the source by calculating the
second moment of the counts weighted by a gaussian with width equal to the
size of the source.  We then require the ellipticity to be less than 0.25.
This does not remove any clusters, but removes some spurious sources due to the
readout streak. 

We also removed cluster candidates where the ratio of counts in the soft band over
the counts in the hard band (the spectral softness, $s$) follows the equation,

$$s-3 \sigma_s^{\frac{1}{2}} \geq 1$$.

This cut cleanly removes some systematic fluctuations in the background.  The
softness of the background is defined as exactly 1, since we set the middle our bands to
be the median of the energy as described previously. For XMM-Newton this results in
energy range of 0.5 to 1.7 keV for the soft band and 1.7 to 7.0 keV for the hard
band.  For Chandra this results in an energy range of 0.5 to 2.2 keV for the
soft band and 2.2 to 7.0 keV for the hard bands.  A high temperature, $\sim 8 keV$, cluster 
has a softness near $2.5$, however, and a lower temperature cluster has a much
higher spectral softness.  

 Finally, we found it necessary for some of the final candidates to be removed by
human inspection, since they were obvious supernovae remnants, planets, or knots in emission
 that had survived all of our cluster selection procedures.  
We utilized a 4 node computing cluster to perform the calculations for this
procedure on all the data sets in about a week, and perfected it after many iterations.  After all the data selection procedures, we have 1198 cluster
candidates as shown in Figure 6.  Many of these clusters were known sources
and were in fact the targets of the relevant observation.  There are 462 clusters that are
outside of 4 arcminutes from the central pointing.  Later we will argue on the
basis of the log N-log S distribution that this is a reliable estimate of the number of new clusters.

After all these cuts we were able to use the AGN in our sample to estimate how many of these
 objects would survive if they were incorrectly chosen as cluster candidates
 due to some mistake in the photon removal procedure.  We find that 19
 AGN would survive these cuts that were more than 4 arcminutes outside the
 central pointing.  Thus, we estimate that the contamination in our survey of
 false clusters is about 3$\%$.  We also need to estimate the efficiency of
 not removing real clusters.  We have, however, performed Monte Carlo
 simulations that found 99$\%$ of all of our simulated clusters.
  We note,
 however, that the cuts we have used above are designed to only remove objects
 that have completely inconsistent properties of a normal cluster.  The one
 cut to worry about in particular might be the size cut, which could remove
 some particularly small groups.  This cut, however, only removes 30$\%$ of
 the candidates so even if some of these were real clusters it probably would not
 be much more than 3 to 5 $\%$.  Furthermore, none of the cuts remove
 any of the brightest 100 already well-known clusters.  Thus, we estimate that
 the efficiency of these candidate cuts is around 95$\%$ and that might
 roughly cancel with the 3$\%$ estimation for contamination.  Both of these
 estimates can be constrained more rigorously with simulations and optical follow-up.

\section{Comparison with previously known clusters}

Since we have used archival observations, a significant fraction of our
sample will be comprised of previously known clusters that were the targets of
their respective observations.
In order to compare the cluster candidates with previously known clusters, we
used the NASA Extragalactic Database (NED).  NED currently contains over
40,000 cluster candidates from over 300 surveys.  In order to check our
cluster candidates reliably, we deconstructed the NED catalog into the
individual cluster catalogs.  We then cross-correlated that catalog with
itself to select objects within 1 arcminute of each other.  A reasonable way
to be confident that the cluster candidates are real is to simply require that
the object appear in multiple catalogs whether they are X-ray surveys,
optical, or any other kind of survey.  We found that 261 (out of 1198 that
passed our selection cuts) clusters in our
survey were previously known.  There was also a significant number of
sources at the center of the field of view that were nearby galaxies and
other extended sources.

\section{Log N-Log S Distribution}

\begin{figure*}[!htb]
  \begin{center}
      \plotone{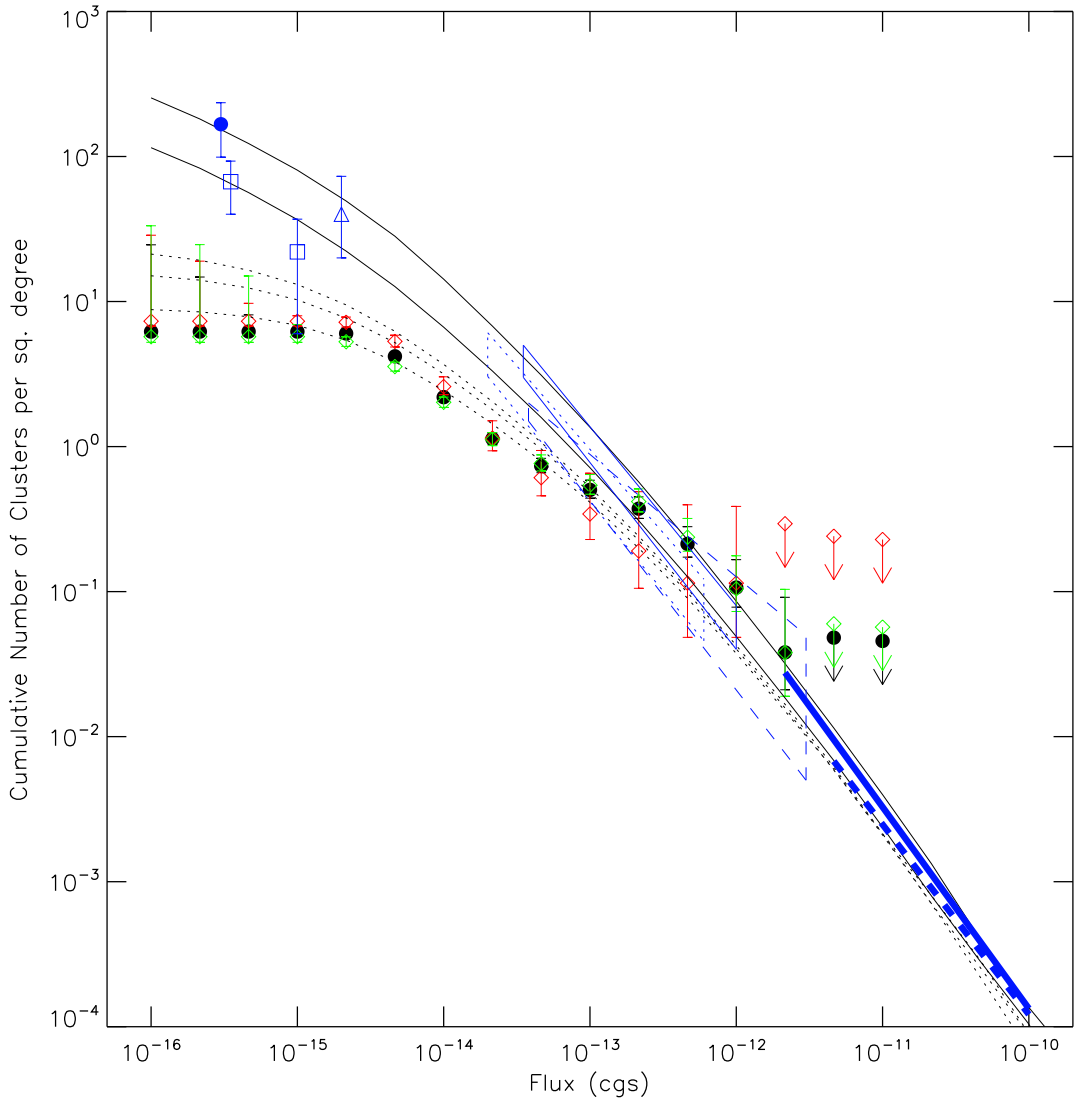}  
  \end{center}
  \caption{\label{fig:lognlogs}  The log N-log S distribution for all the cluster
  candidates (black), the XMM-Newton/Chandra overlap candidates (red diamonds),
  and the Chandra only candidates (green diamonds).
  The models are (black lines from top to bottom): 
  no evolution with $\alpha=1.7$, $L_x^{*min}= 2 \times 10^{41}~\mbox{ergs}~
  \mbox{s}^{-1}$,  $\phi^* =2.7 \times 10^{-7} \mbox{Mpc}^{-3}$, 
   no evolution with $\alpha=1.6$, $L_x^{*min}= 2 \times 10^{42}$,  $\phi^* =1.8 \times 10^{-7}$, and evolution
  models with $z_0=0.7,~1.0,~1.3$ using the latter set of parameters (see text).  The blue lines are those of
  \citealt{bauer} (circle), \citealt{giacconi} (square),
  \citealt{mccardy} (triangle), \citealt{henry} (dashed box), \citealt{rosati2}
  (dotted box), \citealt{vikhlinin} (solid box), \citealt{boehringer2} (solid line
  on right), and \citealt{ebeling} (dotted line on right).}
\end{figure*}

\begin{figure*}[!htb]
  \begin{center}
    \plotone{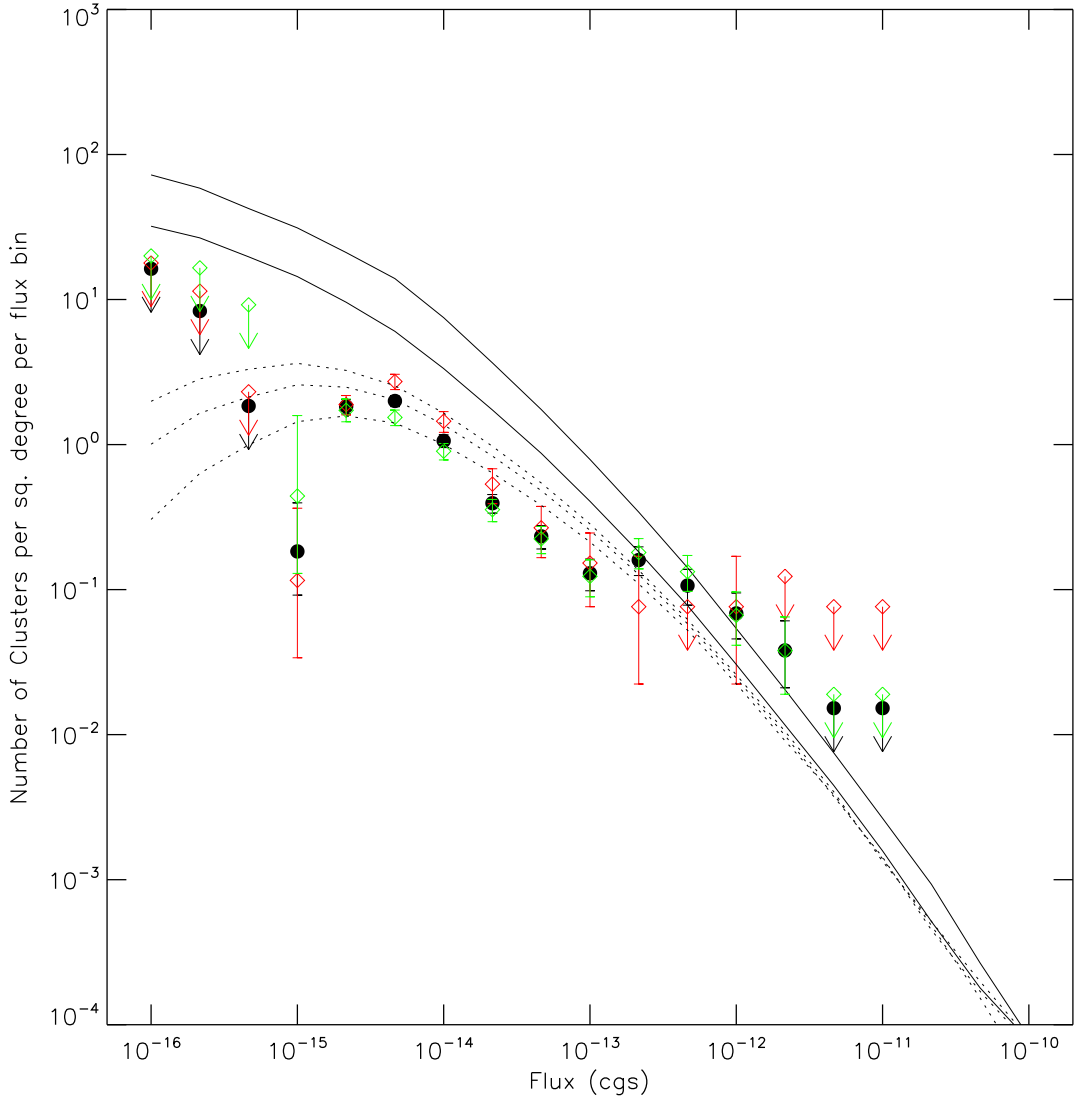}
    \end{center}
\caption{Identical plot to the previous plot except in differential form.}
\end{figure*}

\begin{figure*}[!htb]
  \begin{center}
      \plotone{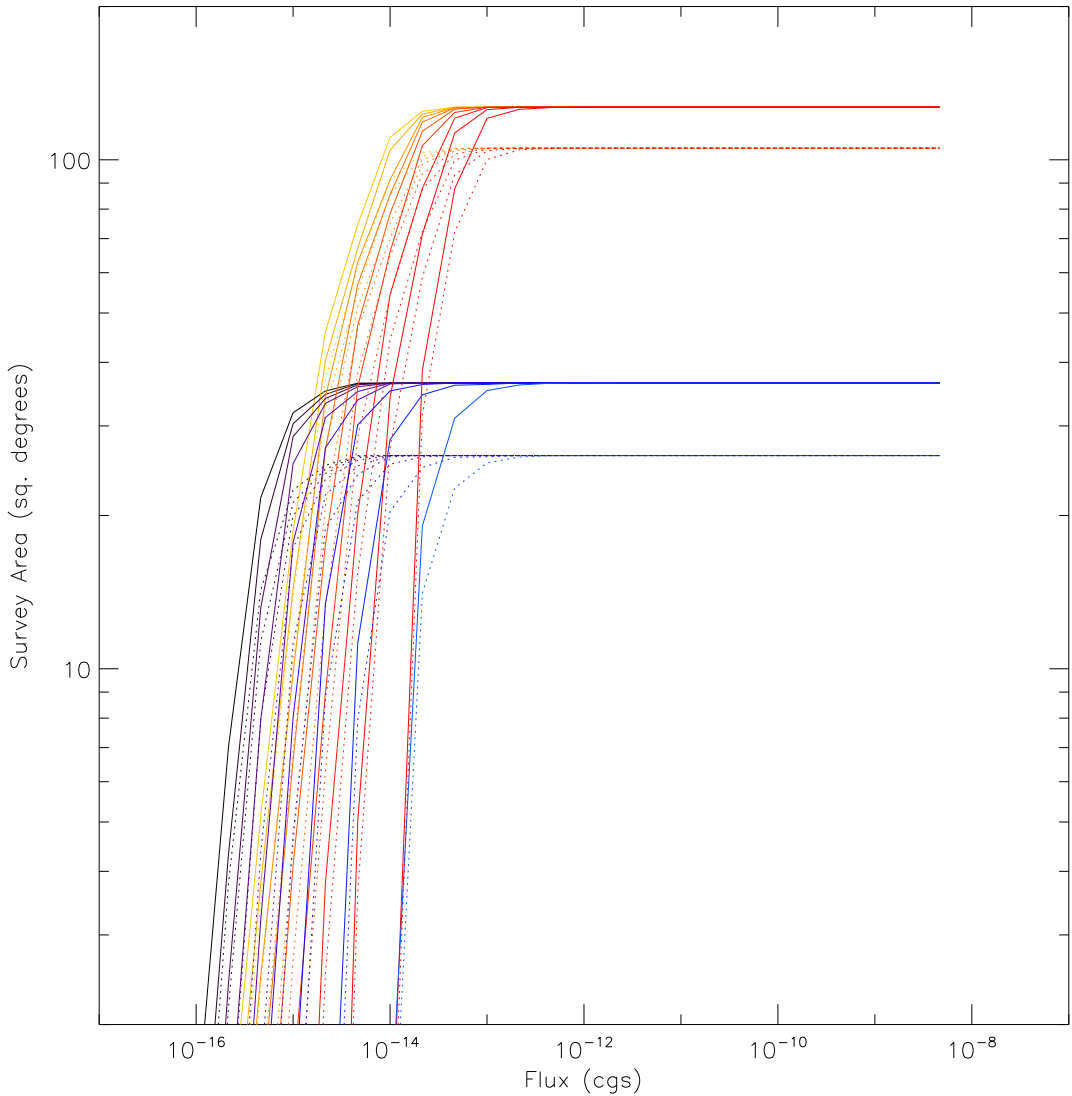}  
  \end{center}
  \caption{\label{fig:surveyarea}  The survey area in square degrees as a
  function of flux for different source sizes (4 (left most) to 256 (right most) arcseconds in
  multiples of 2).  The blue curves are the overlap survey and the red curves
  are the part of the survey where there is only Chandra data.  The
  Chandra-only survey has a larger area, but does not go as deep as the
  XMM/Chandra overlap survey.  The dotted curve shows the area after the 4
  arcminute central pointing region is excluded.}
\end{figure*}

\begin{figure*}[!htb]
  \begin{center}
      \plotone{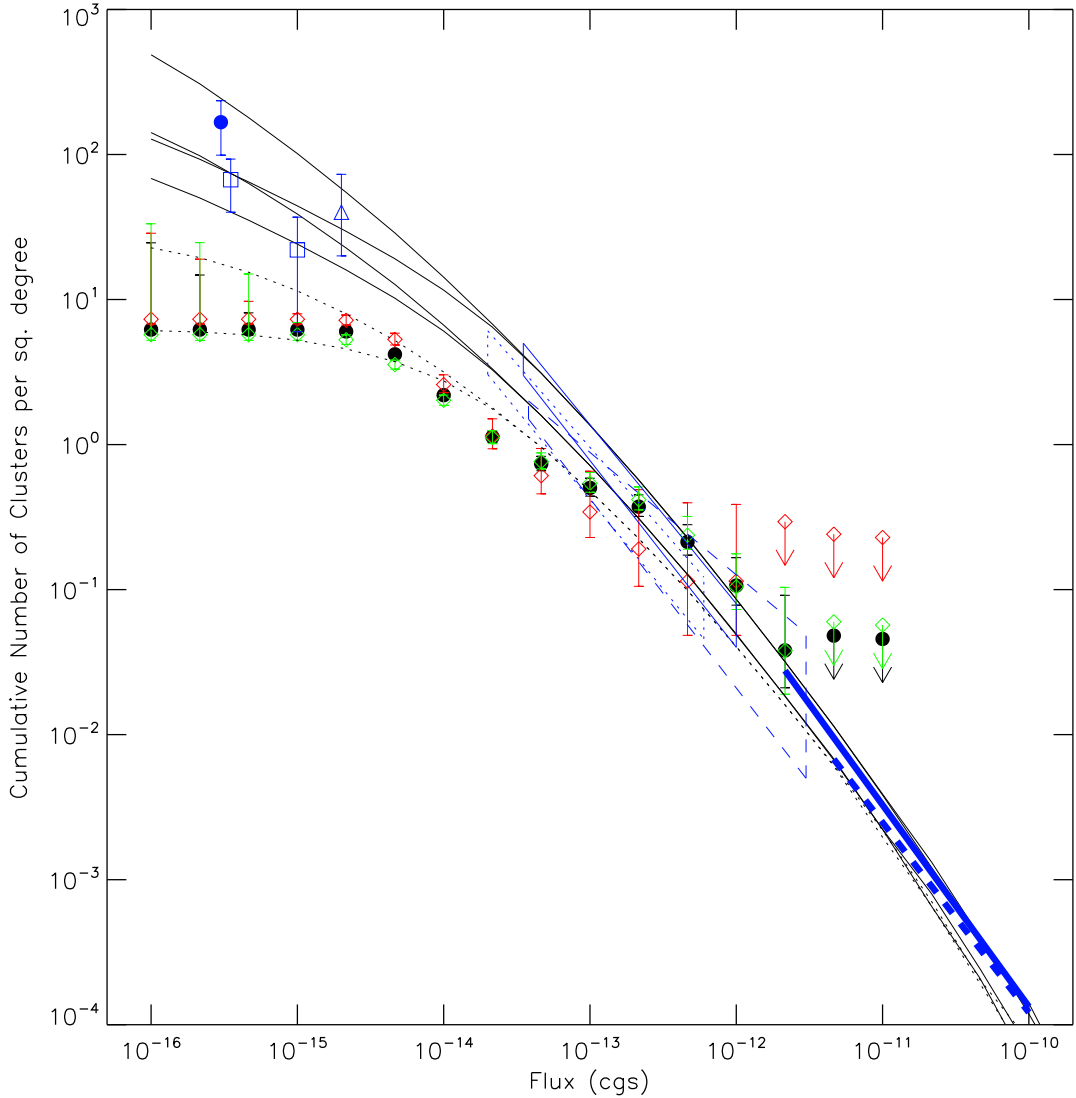}  
  \end{center}
  \caption{\label{fig:lognlogs_syst}  The log N-log S distribution for the cluster
  candidates showing the effect of the size cut.  We plot the same curves as
  in Figure 11, but we show both a size cut of zero arcseconds (lower
  curve) and eight arcseconds for the same model parameters as before except
  we only include $z_0=1.0$ for clarity.  Clearly, the size cut has very
  little effect above fluxes of $10^{-14}~\mbox{ergs}~\mbox{cm}^{-2}~\mbox{s}^{-1}$.}
\end{figure*}

\begin{figure*}[!htb]
  \begin{center}
      \plotone{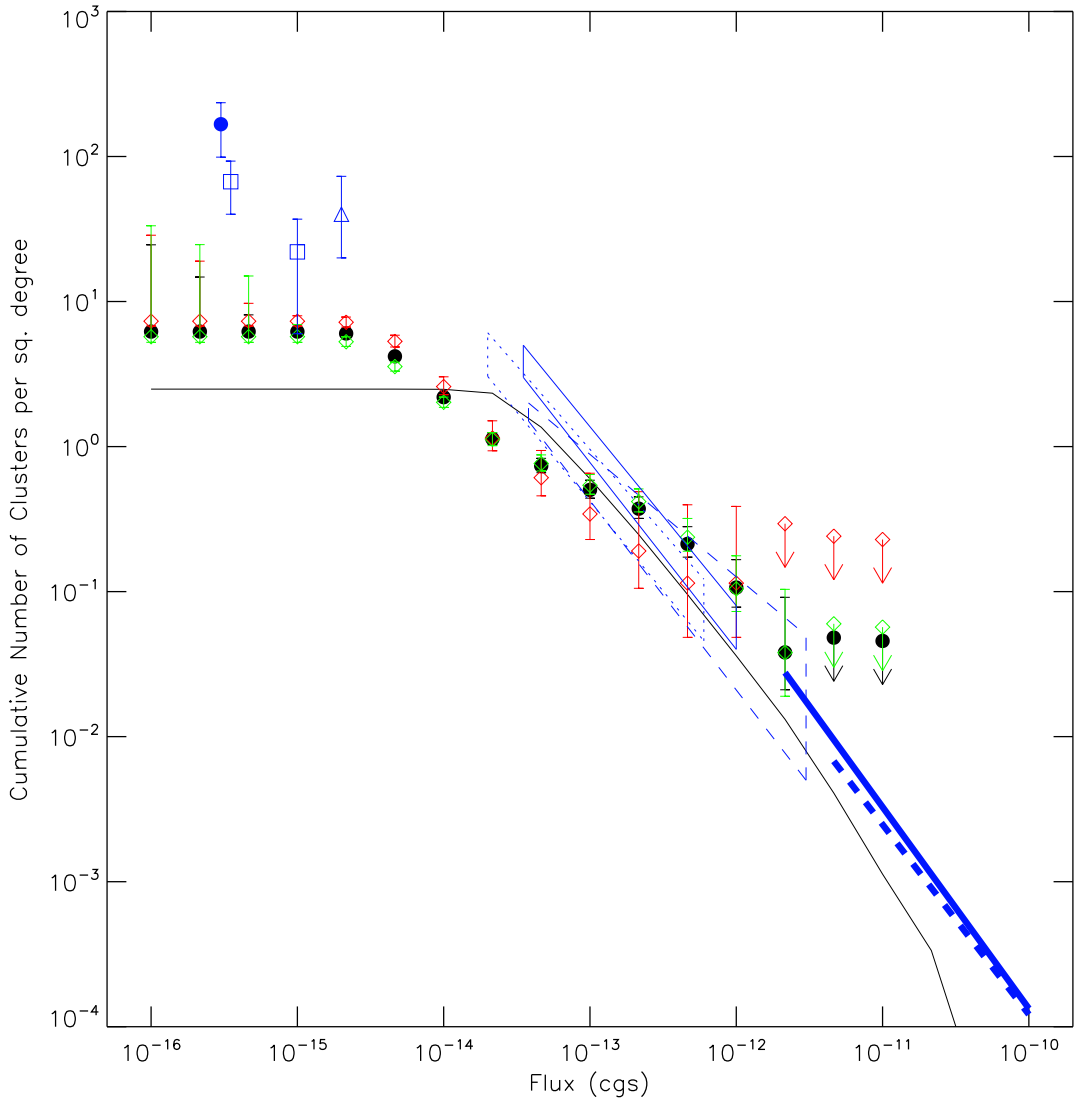}  
  \end{center}
  \caption{\label{fig:structureform}  The expected number density of clusters
  compared to the present number density.  The blue dashed region is the
  region enclosed by our measured exponential function.  The three curves
  correspond to the number density for cluster of mass of $3 \times 10^{13}$,
  $10^{14}$, and $3 \times 10^{14}$ solar masses (from right to left) using
  the calculation described in the text from structure formation theories for
  concordance cosmology.
  The exact shape of the rapid turnover is not important, but these curves
  demonstrate that structure formation theories predict a similar rapid
  evolution in the number density close to the observations.}
\end{figure*}

\begin{figure*}[!htb]
  \begin{center}
      \plotone{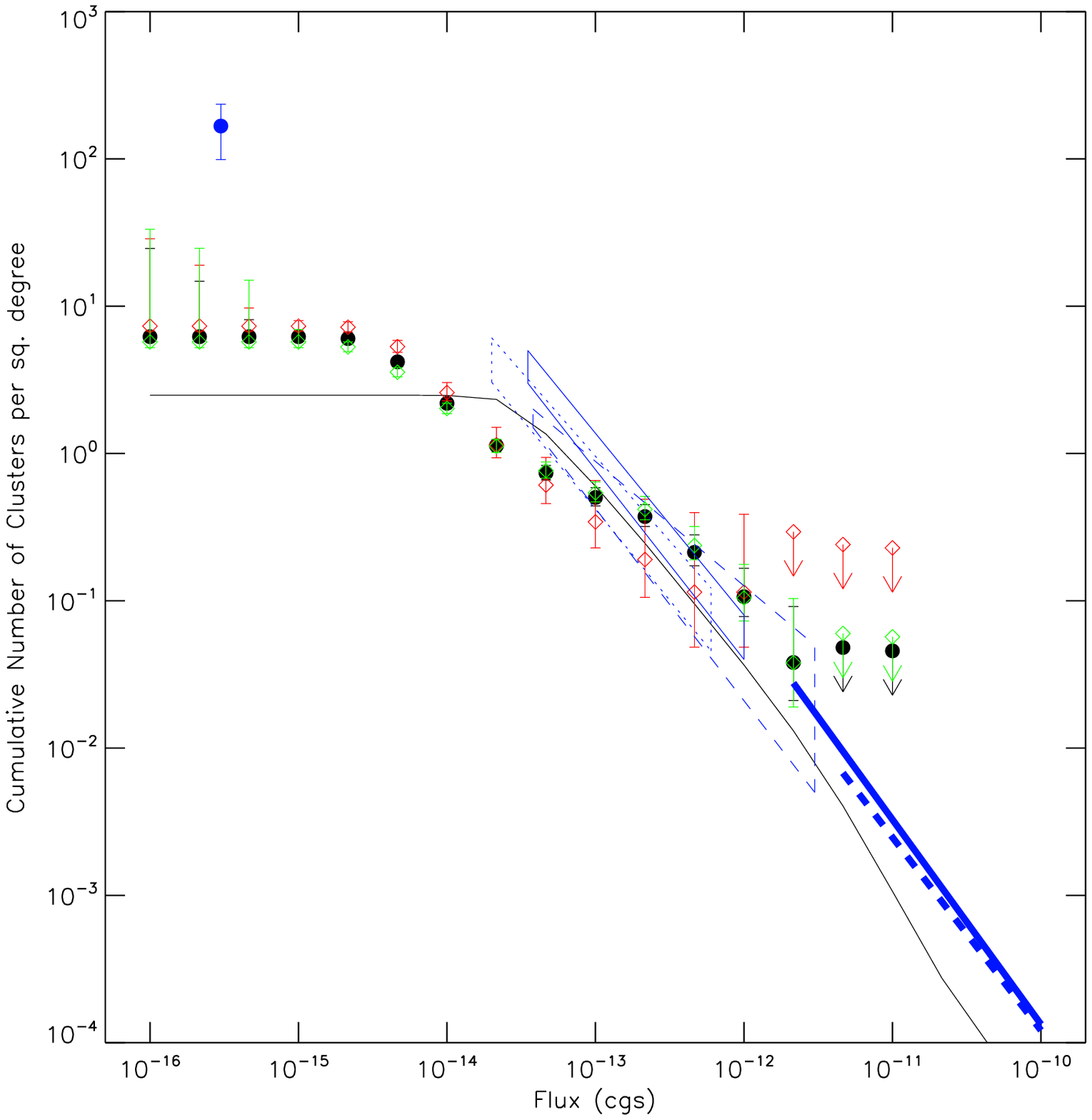}  
  \end{center}
  \caption{\label{fig:lognlogs_j}  The log N-log S distribution as before, but
  using the direct calculation of the distribution from structure formation
  theory.  See the text for further details.}
\end{figure*}

We constructed the log N-log S distribution for the clusters in our sample as
shown in Figure 11.  We plotted the log N-log S distribution in both the
cumulative and differential forms (Figure 12).  The conversion from counts to flux was
calculated by using WebPIMMS using a column density of $2\times10^{20}
\mbox{cm}^{-2}$, $kT=4 keV$, and abundance of 0.4 solar.  Although there is
obviously variation in our clusters, the variation will be less than our flux
bin width and these are expected typical values.
The log N-log S distributions are then calculated
by adding up the number of sources divided by the area of the survey that
source in which that source could have been detected.
Since we have a different flux limit for every point in
the survey, we had to take this into account.  Furthermore, our source
detection sensitivity depends on the object's measured wavelet size.  Therefore, we calculated
the survey area as a function of flux for each object taking into account our
source selection function in Figure 13.  For each square arcminute of the survey we
calculate the total exposure and multiply that by a given flux.  This gives us
an estimate of the number of photons that would be produced by a source at
that particular place in the sky.  Then we see if that would pass both our
significance cut described in Section 3.4 and our signal to noise cut for the background
at that particular place in the field. 
Thus, at high source fluxes the survey area is the full
survey, but at low source flux it will be lower. 
 We have assumed that our other selection cuts do not remove significant
 numbers of real clusters and largely remove spurious data artifacts.  Our
 earlier Monte Carlo simulations indicate for a wide range of sources that
 most of these (99$\%$) are recovered by the source finding algorithm.

 We examined the subset of sources located a distance $A$ from the centers of
 observations in which they were detected, and varied $A$ until the log N-log S
 distribution convered.  This occurs for $A$ $\sim$ 4 arcminutes.  We take
 this as an indication that the application of that cut removes the bias
 associated with selected pointings at known clusters.  We also separated the Log N-Log S points for the overlapping Chandra and
XMM-Newton survey (red diamonds) and the Chandra only data (green diamonds
).  The difference in these surveys gives us a rough measure of the
systematic error, since these are completely different samples.

We modelled the log N-log S distribution using the following method.  First to
be consistent with previous large angle surveys we adopted the Schechter
function as outlined in the review by \citealt{rosati}, 

$$ \frac{dN}{dL_X} = \frac{\phi^{*}}{L_x^*} {\left( \frac{L_x}{L_x^*}
  \right)}^{-\alpha} e^{\left( -L_x/L_x^* \right)} e^{\left(
  -L_x^{*min}/L_x \right)}$$

where we added the last factor to possibly cut off the luminosity on the low
luminosity end to account for small groups that will not even emit X-rays even
if they obey scaling relations.  We used the values of $\alpha=1.7$ (has
15$\%$ scatter in the literature), $L_x^*= 4 \times 10^{44}~\mbox{ergs}~\mbox{s}^{-1}$, $\phi^* =
2.7 \times 10^{-7} \mbox{Mpc}^{-3}$
(has 50$\%$ scatter in the literature) from \citealt{rosati} which summarize the previous work at
the high flux end.  We also used slightly lower values of $\alpha=1.6$ and
$\phi^* = 1.8 \times 10^{-7} \mbox{Mpc}^{-3}$
 for a second calculation than the mean values measured
in other work but still within the observational uncertainties to be
conservative about the measurement of evolution as we discuss below.
We then use this luminosity function and the comoving volume element of the Universe for
$\Lambda CDM$ ($\Omega_m=0.25, \Omega_{\Lambda}=0.75$) to populate the Universe
with clusters as a function of redshift.  Then, we convert the luminosities of the
clusters using the appropriate luminosity distance for the same cosmology.  We
also predict the fraction of these clusters that would be missed by our size
cut by estimating the physical size, $R$ of the cluster following the relation of

$$ R = 1 \mbox{Mpc} {\left( \frac{L_x }{10^{46} \mbox{ergs}~{s}^{-1}} \right)}^{0.45}$$.

\noindent
We determined this relation by looking at the clusters in our sample where we
knew the redshift.  This physical distance is converted into an angular size
using the angular diameter distance for the same cosmologies.  This allows us
to predict the log N-log S distributions in Figure 11.

We varied $L_x^{*min}$ between $2 \times 10^{41}~\mbox{ergs}~\mbox{s}^{-1}$ and
$2 \times 10^{42} \mbox{ergs}~\mbox{s}^{-1}$ which had an minor effect on the
curves.  A luminosity of $2 \times 10^{42} \mbox{ergs}~\mbox{s}^{-1}$ would
correspond to a object with virial temperature around 400 eV and a luminosity
of $2 \times 10^{41}~\mbox{ergs}~\mbox{s}^{-1}$ would correspond with a virial
temperature around 100 eV.  Even if there are many objects in the Universe
with these sizes they would stop emitting significant X-rays somewhere in this
range since the flux is located out of the X-ray band in this temperature range.

Similarly, the cosmological parameters used in the calculation of the volume
element have a small effect as well.  We adopted the concordance values of
$\Omega_m=0.25$, $\Omega_{\Lambda}=0.75$, and $H_0=72 \mbox{km} \mbox{s}^{-1}
\mbox{Mpc}^{-1}$.
 If we assume no evolution in this
luminosity function then the data are completely inconsistent with the model
using the nominal set of parameters for $\phi^{*}$, $L_x^{*min}$, and $\alpha$
as well as a conservative set of values for these parameters as shown in
Figure 11.  On the other hand, if we modify $\phi^{*}$ by a factor of $(1+z)^q$ we can
match the data only if q is greater than -2.  This demonstrates that we have measured the onset of cluster
formation and have observed the dearth of clusters at high redshift.  Since
the $1+z$ formulation shows rapid evolution, we have also simply tried to
represent this rapid evolution as a $e^{ -\frac{z}{z_{0}} }$ instead.  In Figure 11,
  we plot values of $z_0$ of 0.7,1.0, and 1.3 along with the no evolution
  case.  Given the parameters of the model we have used, the data points
  towards rapid evolution at relatively low redshift of z=0.7 to 1.3.  The
  exact value may depend somewhat on our choice of parameters for the
  normalization and slope of the luminosity function.  It also might be
  affected by unaccounted for inefficiencies of the survey which could reduce
  the  number of observed clusters by 10 to 20\%.  Note, however, that the
  size cut only mildly reduces the number of clusters and that cut would have
  to be falsely removing clusters as large as 20 arcseconds instead of 4
  arcseconds to explain the evolution.  To demonstrate this, we show the
  effect of a zero arcsecond and eight arcsecond cut in Figure 13.  Generally,
  there is little effect below a flux of $10^{-14}~\mbox{ergs}~\mbox{cm}^{-2}~\mbox{s}^{-1}$.  For that reason, it is very likely
  that the observed deficit at low fluxes is due to an actual absence of
  clusters at high redshift.  Note also that our exponential function
  describes the number density of clusters and not the absolute number of
  clusters which will peak at higher redshifts.

This rapid evolution is
generically predicted in models of structure formation with a non-negligible
amount of matter in the Universe.  Structure formation models predict a decrease in both the normalization of the cluster mass
function at high redshift as well as a change in shape of the function
\citep{jenkins}.  Furthermore, if there is significant cluster evolution as
we have argued, then the parameterization of evolution in terms of
$(1+z)^{q}$ or $e^{-\frac{z}{z_{0}}}$
is an oversimplification.  In Figure 14, we plot the range of exponential
functions derived to match the log N-log S distribution.  We then overlay the
expected number density of clusters combining the \citealt{jenkins} theory with
the power spectrum of \citealt{eisenstein} in an identical calculation to that of
\citealt{haiman} for three different mass of clusters.  We use the concordance
parameters above and also set $\sigma_8=0.75$, $\Omega_b=0.04$, $w=-1$, and
$n_s=1.0$.  The agreement clearly
demonstrates that the rapid evolution that we have measured is similar to what
is expected for reasonable cluster masses.  Future detailed measurements and modelling when masses and
redshifts can be estimated may measure the exact nature of the cluster
evolution. 

 The structure formation model predicts an evolution that varies as
a function of mass, so we can predict the log N-log S
distribution directly as an additional check.  We must make the additional
assumption of a mass-luminosity
relation and assume that the theoretical mass function agrees well with the
local luminosity function.  We adopt the mass-luminosity relation of
\citealt{stanek} where $L_{44} = 3.8 \left( \frac{M}{10^15 h^{-1}}
\right)^{1.59}$.  We ignored any redshift dependence of the relation to be
conservative, since the change in the relation would make the evolution more
significant. 
We also assume that simulated clusters with calculated
luminosities below $2 \times 10^{42}~\mbox{ergs}~\mbox{s}^{-1}$ are suppressed
by the same exponential function as before.  The direct log N-log S
calculation is shown in Figure 15, and agrees quite well given the
uncertainties in our mass calibration. 

Our results are generally consistent with
  intermediate flux ROSAT surveys at $10^{-13}
  \mbox{ergs}~\mbox{cm}^2~\mbox{s}^{-1}$ as measured in \citealt{henry}, \citealt{rosati2}
  and \citealt{vikhlinin}.  Near a flux of $10^{-14}~\mbox{ergs}~\mbox{cm}^2~\mbox{s}^{-1}$ our results deviate somewhat
  particularly with those of \citealt{vikhlinin} and are closest to those of \citealt{henry}.  A combination of some
  non-cluster contamination and an underestimate of the survey sensitivity
  could explain these discrepancies.  Considering the error ranges, however, these
  measurements could still be consistent with some mild cluster evolution.  Our
  results are complementary to those at high fluxes,
  e.g. \citealt{boehringer2}, and we have fixed our theoretical model
  consistent with those results.

Our results are also broadly consistent with the recent work of Vikhlinin
where cluster masses have been estimated for a sample of 36 clusters
\citealt{vikhlinin2}.  In this work, he demonstrated that a lower redshift
sample ($z\sim0.1$) had a different and lower normalization mass function than
the cluster at higher redshift $z\sim0.5$.  The results were consistent with
the concordance structure formation model discussed above.  In a later work,
Vikhlinin measured the equation of state parameter for dark energy to 10$\%$ \citealt{vikhlinin3}.
Since our sample is larger and deeper, it might be possible to measure the
equation of state parameter to 2$\%$ in a different mass and redshift regime.

Our results are inconsistent, however, with the log N-log S distribution determined
using the Chandra deep field \citep{bauer}.  This was determined by using
  130 square arcminutes and finding six candidates clusters in that survey
  that suggested there was no evolution in the cluster luminosity function.
  These objects, however, have signal to noise less than 1 (table 1, column 6
  of their paper), so would not survive
  our signal to noise cut.  If we remove our signal to noise cut we find 3
  out of the six objects.  If only two or three are clusters, then we would be
  consistent with this measurement.  Similarly, both \citealt{mccardy} and
  \citealt{giacconi} found clusters in deep and narrow surveys.
  \citealt{giacconi} do not make a claim about cluster number density.  If we
  convert their data to log N-log S as in \citealt{rosati}, then they are
  consistent with our measurements at $10^{-15}~\mbox{ergs}~\mbox{s}^{-1}$ but inconsistent at lower fluxes.
  \citealt{giacconi} argue, however, that many of these sources could be
  contaminated by hard X-ray emission from AGN. 

  Our results are consistent with
previous measurements of the evolution of the luminosity functions where the
redshifts were known in the Rosat Deep Cluster Survey  and the Einstein Medium
Sensitivity Surveys (\citealt{gioia90} and see Figure 9 in \citealt{rosati}),
although the interpretation is somewhat different.  There the authors considered both an evolution in the
normalization and the value of $L_X^{*}$ (in their notation the parameters are
A and B).  In our case, the evolution factor, $q$, is equivalent to $A+B$.  The
previous results have constrained the value of $A+B$ to be approximately
equal to $-3$, which is also completely consistent with our result.  The other authors
have argued, however, that the data are more consistent with mostly a non-zero
value of $B$, which we cannot distinguish from the flux distribution alone.  In addition, our
sample is much deeper and it is likely that the Schechter formulation is not
even correct to such a large range in redshifts.  In any case, negative
evolution in the clusters x-ray luminosity function and the emergence of clusters in
the Universe has been clearly observed.  Additional detailed photon Monte Carlo
simulations like those we have proposed above as well as optical follow-up to
remove spurious sources are needed to verify the efficiency and contamination
we have claimed in the log N-log S calculation.  However, the large deviation
of the Log N-Log S curve at
$10^{-14}~\mbox{ergs}~\mbox{cm}^{-2}~\mbox{s}^{-1}$ compared to the future
model is beyond any reasonable expected systematic error.

\section{Future Work}

There are several research areas that can be pursued in the future with this
cluster sample.  First, the efficiency of the survey can be carefully
calculated by futher Monte Carlo simulations.  
Second, the bias of the survey can be
more accurately estimated by performing a blind survey on a modest field.
Third, the redshifts of the new clusters can be measured by optical
and IR follow-up including cluster verification.
Fourth, the cluster abundance can be measured by combining the
efficiency calculation and the redshifts.  Fifth, the X-ray estimated cluster
masses can be compared to weak lensing measurements for at least some of the
clusters.  Then by assigning these clusters to a given mass and redshift, the
structure formation theory can be tested in great detail.
Finally, the XMM-Newton and Chandra archive continues to accumulate
more data, so a much larger survey can be performed in the future.
Some portion of this additional work is necessary prior to a robust
estimate systematic errors of cosmological parameters.

\acknowledgements

We acknowledge many helpful conversations with Karl Andersson, Kari Frank,
Suzanne Nichols, Frits Paerels, Masao Sako, Caleb Scharf, and Mark Voit.
Support for this work for JRP, JGJ, and JB is from a NASA grant NNX07AH51G from the
NASA-ADP program.  JRP is also supported from a grant from the Purdue Research Foundation.

\end{document}